\documentclass[11pt]{cernrepdbg}
\usepackage{graphicx}
\usepackage{here}
\usepackage{cite}
\usepackage{amsmath}
\usepackage{amssymb}
\usepackage{epsfig}
\usepackage{axodraw}
\usepackage{rotating}  

\newcommand{\href}[2]{ \texttt{#1} }

\graphicspath{{figures/}}
\DeclareGraphicsExtensions{.eps,.eps.gz,.ps,.ps.gz} 

\newcommand{\txt}[1]{{\mathrm{#1}}} 

\newcommand{\Matrix}{\mbox{${\cal M}$}}

\newcommand{\MSbar}{\overline{\mathrm{MS}}}

\newcommand{\PYTHIA}{{\tt PYTHIA}}

\newcommand{\ppORppbar}{pp\!\!\!\!^{^{(-)}}}
\newcommand{\qORqbar}{q\!\!\!\!^{^{(-)}}}

{ \begin{enumerate} \setlength{\itemsep}{0pt} }%
{ \end{enumerate} }

{ \begin{sideways} \begin{minipage}[t]{\textheight} }%
{ \end{minipage} \end{sideways} }

\newenvironment{fminipage}%
{\begin{Sbox}\begin{minipage}}%
{\end{minipage}\end{Sbox}\fbox{\TheSbox}}
{\begin{center}\begin{fminipage}}%
{\end{fminipage}\end{center}}

\newenvironment{ovalminipage}%
{\begin{Sbox}\begin{minipage}}%
{\end{minipage}\end{Sbox}\ovalbox{\TheSbox}}

{\begin{center}\begin{ovalminipage}}%
{\end{ovalminipage}\end{center}}

\newcommand{\LineFilledHeading}[1]{
        \begin{center}%
        \hrulefill \raisebox{-1mm}{ {\bf #1} } \hrulefill%
        \end{center}
}

\newcommand{\programAbstract}[2]{
    \LineFilledHeading{#1}
    {\small\em (Contributed by: #2)}\newline }

\newcommand{\programInfo}[3]{ \noindent 
     {\small\underline{Authors:}} #1 \\ {\small\underline{Ref:}} #2 \\
     {\small\underline{Webpage:}} \href{#3}{#3} \\ }
\newcommand{\programVersion}[1]{ \noindent 
     {\small\underline{Current Version:}} #1 \\ }

\newcommand{\programsubsection}[1]{\vspace{0.1cm}
\noindent \underline{{\bf #1}} }

\def\VEV#1{\left\langle #1\right\rangle}

\def\beq{\begin{equation}}
\def\eeq{\end{equation}}

\def\beqn{\begin{eqnarray}}
\def\eeqn{\end{eqnarray}}
\relax

\newcommand\sss{\scriptscriptstyle}

\newcommand\as{\alpha_{\sss S}}         
\newcommand\ep{\epsilon}

\newcommand\Hfun{{\cal H}}

\newcommand\MSB{{\rm \overline{MS}}}

\newcommand\CF{C_{\sss F}}

\newcommand\pt{p_{\sss T}}
\newcommand\kt{k_{\sss T}}

\newcommand\epem{e^+e^-}

\newcommand\xsec{\frac{d\sigma}{dx}}

\newcommand\xsborn{\left(\xsec\right)_{\sss B}}
\newcommand\xsvirt{\left(\xsec\right)_{\sss V}}
\newcommand\xsreal{\left(\xsec\right)_{\sss R}}
\newcommand\stepf{\Theta}
\newcommand\xc{x_c}

\newcommand\wEV{w_{\sss EV}}
\newcommand\wCNT{w_{\sss CT}}

\begin{document}
\bibliographystyle{unsrt}

\title{Les Houches Guidebook to Monte Carlo Generators\\ for Hadron 
       Collider Physics}
\date{\today}
\author{
\underline{Editors:} 
M.A.~Dobbs$^{1}$, 
S.~Frixione$^{2}$, 
E.~Laenen$^{3}$, 
K.~Tollefson$^{4}$
\\ \underline{Contributing Authors:}
H.~Baer$^5$,
E.~Boos$^6$,
B.~Cox$^7$,
M.A.~Dobbs$^1$,
R.~Engel$^{8}$,
S.~Frixione$^2$,
W.~Giele$^{9}$,
J.~Huston$^{4}$,
S.~Ilyin$^6$,
B.~Kersevan$^{10}$,
F.~Krauss$^{11}$,
Y.~Kurihara$^{12}$,
E.~Laenen$^3$,
L.~L\"onnblad$^{13}$,
F.~Maltoni$^{14}$,
M.~Mangano$^{15}$,
S.~Odaka$^{12}$,
P.~Richardson$^{16}$,
A.~Ryd$^{17}$,
T.~Sj\"ostrand$^{13}$,
P.~Skands$^{13}$,
Z.~Was$^{18}$,
B.R.~Webber$^{19}$,
D.~Zeppenfeld$^{20}$
}

\institute{\centering{\small
\vspace{0.5cm}
$^1$Lawrence Berkeley National Laboratory, Berkeley, CA 94720, USA \\
$^2$INFN, Sezione di Genova, Via Dodecaneso 33, 16146 Genova, Italy\\
$^3$NIKHEF Theory Group, Kruislaan 409, 1098 SJ Amsterdam, The Netherlands\\
$^4$Department of Physics and Astronomy, Michigan State University,
      East Lansing, MI 48824-1116, USA\\
$^5$Department of Physics, Florida State University, 511 Keen Building,
Tallahassee, FL 32306-4350, USA \\
$^{6}$Moscow State University, Moscow, Russia\\
$^{7}$Dept of Physics and Astronomy, University of Manchester, Oxford Road, 
Manchester, M13 9PL, U.K. \\
$^{8}$ Institut f\"{u}r Kernphysik,
Forschungszentrum Karlsruhe, Postfach 3640, D
- 76021 Karlsruhe, Germany \\
$^{9}$Fermi National Accelerator Laboratory, Batavia, IL 60510-500, USA\\
$^{10}$Jozef Stefan Institute, Jamova 39, SI-1000 Ljubljana,Slovenia;
Faculty of Mathematics and Physics, University of Ljubljana,
Jadranska 19,SI-1000 Ljubljana, Slovenia \\
$^{11}$Institut f{\"u}r Theoretische Physik, TU Dresden, 01062 Dresden, 
 Germany\\
$^{12}$KEK, Oho 1-1, Tsukuba, Ibaraki 305-0801, Japan\\
$^{13}$Department of Theoretical Physics, Lund University, S-223 62
Lund, Sweden \\
$^{14}$Centro Studi e Ricerche ``Enrico Fermi'', via Panisperna, 89/A
- 00184 Rome, Italy\\
$^{15}$CERN, CH--1211 Geneva 23, Switzerland\\
$^{16}$Institute for Particle Physics Phenomenology, University of Durham, 
 DH1 3LE, U.K.\\
$^{17}$Caltech, 1200 E. California Bl., Pasadena CA 91125, USA \\
$^{18}$Institute of Nuclear Physics PAS, 31-342 Krakow,
ul.~Radzikowskiego 152, Poland \\
$^{19}$Cavendish Laboratory, Madingley Road, Cambridge CB3 0HE, U.K. \\
$^{20}$Department of Physics, University of Wisconsin, Madison, WI 53706, USA\\
}}

\maketitle
\begin{abstract}
Recently the collider physics community has seen significant advances
in the formalisms and implementations of event generators. 
This review is a primer of the methods commonly used
for the simulation of high energy physics events at particle
colliders. We provide brief descriptions, references, and links to the
specific computer codes which implement the methods.
The aim is to provide an overview of the available tools, allowing the
reader to ascertain which tool is best for a particular application,
but also making clear the limitations of each tool.
\end{abstract}

\begin{center}
{\it Compiled by the Working Group on Quantum ChromoDynamics and the
  Standard Model for the Workshop ``Physics at TeV Colliders'', Les
  Houches, France, May 2003.} \\
{\bf \today}
\end{center}

\clearpage
\tableofcontents
\setcounter{footnote}{0}

\section[Introduction]{INTRODUCTION~\protect\footnote{Contributed by: 
the editors.}}\label{s_introduction}

The complexity and number of simulation programs for hadron
colliders has grown considerably with the prospects of LHC physics
approaching and Tevatron Run II results coming in. With these programs has
come a shift towards increased modularity. A physicist analysing
hadron collider data often obtains the most accurate theoretical
predictions by combining components of many different simulation
programs---minimum bias from one generator, the signal process from
another, and yet more programs for background generation. This sort of
diversification is also happening for the generation of a single
process. It is becoming feasible to use one program to produce a hard
process, another to evolve the event through a parton shower
algorithm, and perhaps a third to hadronize the coloured products of
the shower. With this sort of modularity, the complexity of Monte
Carlo simulation tools is reaching that of a complicated detector
system. At the same time the expertise needed by the users is
increasing.  At the very start of a physics analysis, the experimenter
is confronted with a simple question, {\it which Monte Carlo tools are
best suited to map the theoretical prediction for my measurement onto
the experimental result?} 

The goal for this guidebook is to provide users inexperienced with
event simulation a starting point to answer the ``which tools?''
question. A complete description of Monte Carlo generator techniques
would require a many-volumed book. Instead we provide the 
basic definitions and explanations which a new reader
will need to appreciate the literature. We do so in the most
politically incorrect way, by not quoting the original papers in most
cases (since the foundations are textbook matter by definition), 
and striving for plain jargon-free language.  We follow this
with abstracts describing many of the currently available simulation
programs, aiming to serve as a jumping off point into the specific
references documenting the programs and the techniques employed within
them. The abstracts will also point users to the (author supplied)
correct references for citations to their papers.

Finally, the editors wish to apologise to the authors of Monte Carlo
codes for which we have not provided abstracts. We chose to
restrict this work to hadron colliders only, and limited the scope to
general purpose techniques, which are more or less directly related
to event generator codes. For this reason, we could not list the many
NLO or resummation programs which are available for specific processes. 
Despite this limitation, there are still a large number of program abstracts
included in this guidebook. In all likelihood we have missed a few packages
and we apologise to those authors in advance.

\section[The Simulation of Hard Processes]
{THE SIMULATION OF HARD PROCESSES~\protect\footnote{Contributed by: 
M.~Dobbs, S.~Frixione.}}\label{s_simulation}

Theoretical predictions form an integral part of any particle
physics experiment. On one hand, they help to design the detectors
and to define the experimental strategies. To serve such a purpose,
these predictions must reproduce as closely as possible the collision
processes taking place in real detectors. A largely successful way of
achieving this goal is through the so-called {\em event generator}
codes, which are used to produce hypothetical events with the 
distribution predicted by theory---i.e.\ the frequency we expect
the events to appear in Nature. On the other hand, for an unambiguous
interpretation of the experimental results (for example, 
extracting with high precision the non-computable parameters of the 
theory or deciding whether some new physics phenomena has been
observed) other types of codes, which we shall call {\em cross
section integrators}, are better suited than event generators.
In a loose sense, these codes can also output events (see 
sect.~\ref{sect:NLOcomp} for a precise definition); however, 
such events can be used only to predict a limited number of 
observables (for example, the transverse momentum of single-inclusive 
jets) and are not a faithful description of actual events taking
place in real detectors.

Currently, event generators and cross section integrators have reached
a considerable sophistication.  The purpose of this introductory section
is to show that both of them originate from the very same simple 
description of an elementary process (denoted as {\em hard subprocess}
henceforth) and not necessarily a physically-observable one.

To stress the latter point, let us design a gedanken experiment
which, at an imaginary accelerator that collides 45~GeV $u$-quarks 
with 45~GeV $\bar{u}$-quarks, observes a $d\bar{d}$ quark pair produced
through the decay of a $Z^0$. The process of interest is therefore
$u\bar{u} \rightarrow Z^0 \rightarrow d\bar{d}$ at 90~GeV. Any 
theoretical model describing this process must start from the
knowledge of its cross section
\begin{equation}\label{e_xsec_uuZ}
d\sigma(u\bar{u}\rightarrow Z^0 \rightarrow d\bar{d}) ~=~ 
\frac{1}{2\hat{s}}\;
|\Matrix(u\bar{u}\rightarrow Z^0 \rightarrow d\bar{d})|^2 \;
\frac{ d\cos\theta d\phi } { 8 (2\pi)^2 }\,,
\end{equation}
where the decay angles ($\theta,~\phi$) of the $Z^0$, are the two 
degrees of freedom of the problem\footnote{The rotational symmetry of the 
collision implies that the differential cross section is independent of the 
azimuthal angle $\phi$.}. $\Matrix$ is the relevant matrix element and 
$\hat{s}$ is the centre-of-mass energy squared.

We can now use eq.~(\ref{e_xsec_uuZ}) to write an event generator or
a cross section integrator. The first step
is to {\it sample the phase space}. The phase space is the
multi-dimensional hypercube which spans all of the degrees of
freedom. For this process it is the two dimensional space
\mbox{$-1<\cos\theta<1,~0<\phi<2\pi$}. The procedure of choosing 
the \mbox{$\cos\theta,~\phi$} variables using a uniformly distributed 
random number generator is said to define a candidate event. The candidate 
event's differential cross section (or {\it event weight}) $d\sigma$ is
calculated from eq.~(\ref{e_xsec_uuZ}) and is directly related to the
probability of this event occurring. The average of many candidate
event weights $\langle d\sigma \rangle$ is an approximation to the
integral $\int d\sigma$ and converges to the measured cross section.

At this point the candidate events are distributed flat in phase
space and there is no physics information in the distributions. Two
methods can be used to derive physical predictions from these
candidate events: (A) the event weights may be used to create
histograms representing physical distributions, or (B) the events may
be {\it unweighted} such that they are distributed according to the
theoretical prediction. Procedure (A) is very simple and is what is
done for cross section integrators. A histogram of some relevant 
distribution (e.g.\ the transverse momentum of the $d$ quark) is
filled with the event weights from a large number of candidate
events. The individual candidate events do {\em not} correspond to
anything observable but, in the limit of an infinite number of candidate 
events, the distribution is exactly the one predicted by 
eq.~(\ref{e_xsec_uuZ}).  Procedure (B) is a bit more involved, has added
advantages, and is what is done in event generators. It produces events with
the frequency predicted by the theory being modelled, and the individual
events represent what might be observed in a trial experiment---in this sense
unweighted events provide a genuine simulation of an experiment.

The {\it hit-and-miss} technique (also known as the
acceptance-rejection method or the Von Neumann method) is normally
used to unweight events. To apply the method, the maximum event weight
$d\sigma_\txt{MAX}$ must be known. For this process, the maximum
occurs when one of the final state quarks is collinear
with one of the initial state quarks, so
it is easy to calculate $d\sigma_\txt{MAX}$ by inserting these
conditions ($\cos\theta = \pm 1$) into eq.~(\ref{e_xsec_uuZ}).  For more
complicated processes the maximum event weight can be approximated by
randomly scanning the parameter space. For each candidate event, the
ratio of event weight over the maximum event weight
$d\sigma/d\sigma_\txt{MAX}$ is compared to a random number $g$
generated uniformly in the interval (0,1). Events for which the ratio
exceeds the random number ($d\sigma/d\sigma_\txt{MAX}>g$) are
accepted; the others are rejected. The accepted events have the
frequency and distribution predicted by eq.~(\ref{e_xsec_uuZ}) and
represent the physical expectation for the imaginary $u\bar{u}$
collider experiment. 

We have now learned the basics of the construction of an event generator
or of a cross section integrator. Unfortunately, the process in
eq.~(\ref{e_xsec_uuZ}) is non physical.  This evident fact can
be stressed in two different ways:
\begin{itemize}
\item[{\em a)}] The kinematics of the process is trivial; the $Z^0$
 has transverse momentum equal to zero.
\item[{\em b)}] Quark beams cannot be prepared and isolated quarks
 cannot be detected.
\end{itemize}
Items {\em a)} and {\em b)} have a common origin.  In eq.~(\ref{e_xsec_uuZ})
the number of both initial- and final-state particles is fixed, i.e. there
is no description of the radiation of any extra particles. This radiation
is expected to play a major role, especially in QCD, given the strength
of the coupling constant.  Let's therefore restrict ourselves, in what 
follows, to the case of QCD; although many of these concepts remain 
valid in the context of the electroweak theory.

In the case of item {\em a)}, the extra radiation taking place on top 
of the hard subprocess corresponds to considering higher-order corrections 
in perturbation theory. In the case of item {\em b)}, it can be viewed as
an effective way of describing the dressing of a bare quark which 
ultimately leads to the formation of the bound states we observe
in Nature ({\em hadronization}). Thus, any event generator or 
cross section integrator which aims at giving a realistic description
of collision processes must include: 
\begin{itemize}
\item[{\em i)}] A way to compute exactly or to estimate the effects of
higher-order corrections in perturbation theory.
\item[{\em ii)}] A way to describe hadronization effects.
\end{itemize}
Different strategies have been devised to solve these problems.  They
can be quickly summarised as follows:

\noindent
\begin{center}
{\bf Higher orders}
\end{center}
\begin{itemize}
\item[{\em i.1)}]  Compute exactly the result of a given (and usually 
 small) number of emissions.
\item[{\em i.2)}]  Estimate the dominant effects due to 
 emissions at all orders in perturbation theory.
\end{itemize}

\noindent
\begin{center}
{\bf Hadronization}
\end{center}
\begin{itemize}
\item[{\em ii.1)}] Use the QCD-improved version of Feynman's parton model 
 ideas ({\em factorization theorem}) to describe the parton $\leftrightarrow$
 hadron transition.
\item[{\em ii.2)}] Use phenomenological models to describe the parton 
 $\leftrightarrow$ hadron transition at mass scales where perturbation
 techniques are not applicable.
\end{itemize}

The simplest way to implement strategy {\em i.1)} is to consider
only those diagrams corresponding to the emission of real particles.
Basically, the number of emissions coincides with the perturbative
order in $\as$. This choice forms the core of {\em Tree Level Matrix
Element} generators, described in sect.~\ref{sect:TLMEG}.  These 
codes can be used either within a cross section integrator or within
an event generator. With currently available techniques, the maximum 
number of emissions is between five and ten. A more involved procedure 
aims at computing all diagrams contributing to a given perturbative order in
$\as$, which implies the necessity of considering virtual emissions
as well as real emissions. Such {\em N$^k$LO computations},  
reviewed in sect.~\ref{sect:NLOcomp}, are technically quite challenging
and satisfactory general solutions are known only for the case of one
extra emission (i.e., NLO). Until recently, these computations have been 
used only in the context of cross section integrators; their use within
event generators is a brand new field (see sect.~\ref{sect:MEwPS}).

Strategy {\em i.2)} is based on the observation that the dominant effects in
certain regions of the phase space have almost trivial dynamics, such that
extra emissions can be recursively described. There are two vastly different
classes of approaches in this context. The first one, called {\em resummation}
(see sect.~\ref{s_resummation}), is based on a procedure which generally works
for one observable at a time and, so far, has only been implemented in cross
section integrators.  The second procedure forms the basis of the {\em Parton
Shower} technique (see sect.~\ref{s_shg}) and is, by construction, the core of
event generators.  This procedure is not observable-specific making it more
flexible than the first approach, but it cannot reach the same level of
accuracy as the first, at least formally.

At variance with the solutions given in items {\em i.1)} and {\em i.2)}, 
solutions to the problem posed by hadronization always involve
some knowledge of quantities which cannot be computed
from first principles (pending the lattice solution of the theory)
and must be extracted from data. The factorization theorems mentioned
in {\em ii.1)} are briefly described in sect.~\ref{sect:NLOcomp} and
are the theoretical framework in which cross section integrators are
defined. Parton shower techniques, on the other hand, are used
to implement strategy {\em ii.2)} (see sect.~\ref{s_shg}) in the
context of event generators.

Each of the strategies outlined above, and the codes implementing
them, have strengths and weaknesses that must be considered in order to
choose the best tool for studying the problem of interest. The following
scheme gives a first, rough classification and points to the sections 
where the characteristics of each approach are described
in more detail:

\begin{itemize}
\item {\em If hadronization is expected to play a major role,} use an event
 generator which incorporates a shower and hadronization mechanism
 (sect.~\ref{s_shg}).
\item {\em If hadronization is not a factor,} then cross section integrators
 are sufficient; tree level (sect.~\ref{sect:TLMEG}), NLO 
 (sect.~\ref{sect:NLOcomp}) or resummed (sect.~\ref{s_resummation})
 computations can be adopted.
\item {\em If the analysis studies the peak of the cross section,} event
 generators (sect.~\ref{s_shg}) or cross section integrators implementing
 resummation (sect.~\ref{s_resummation}) should be used.
\item {\em If the analysis studies the tail of the cross section,} then 
multi-leg, tree-level (sect.~\ref{sect:TLMEG}) and NLO 
(sect.~\ref{sect:NLOcomp}) results are usually necessary.
\end{itemize}

Clearly, one should aim for the optimal tool which is able to give correct
predictions both at the peak and in the tail of the cross section. 
Nowadays, this is not just an academic exercise because most of the
analyses performed at the Tevatron and especially at the LHC demand
the construction of such a tool. There has been considerable progress
in the past few years in this direction since we have basically learned
how to merge the techniques for fixed-order matrix element computations
with those relevant to parton shower simulations. More details will
be given in sect.~\ref{sect:TLMEG} and sect.~\ref{sect:MEwPS}.

\section[Tree Level Matrix Element Generators]
{TREE LEVEL MATRIX ELEMENT GENERATORS~\protect\footnote{Contributed by: 
M.~Dobbs, S.~Frixione.}}\label{sect:TLMEG}

In this section we describe codes which allow the computation of 
tree-level matrix elements with a fixed number of legs
(i.e.\ fixed number of partons in the final state). These parton-level
generators describe a specific final state to lowest order in
perturbation theory--virtual loops are {\em not} included in the 
matrix elements. This implies that all complications involving
the regularization of matrix elements are avoided, and the codes
are based either on the direct computation of the relevant Feynman
diagrams or on the solutions of the underlying classical field theory.
We shall not describe these computational techniques in this review;
the interested reader will find the appropriate literature cited
in the papers representing the codes listed below.

These programs generally do not include any form of hadronization,
thus the final states consist of bare quarks and gluons. 
The kinematics of all hard objects in the event are explicitly
represented and it is simply assumed that there is a one to one
correspondence between hard partons and jets.

However, this assumption may cause problems when interfacing these codes 
to showering and hadronization programs such as HERWIG or PYTHIA; a step 
which is necessary in order to obtain a physically sensible description of the 
production process. In fact, a kinematic configuration with $n$ final-state
partons can be obtained starting from $n-m$ partons generated by the
tree-level matrix element generator, with the extra $m$ partons provided 
by the shower. This implies that, although the latter partons are generally 
softer than or collinear to the former, there is always a non-zero probability
that the {\em same} $n$-jet configuration be generated starting from 
{\em different} $(n-m)$-parton configurations. In other words, since 
tree-level matrix elements do have soft and collinear singularities (see 
sect.~\ref{sect:NLOcomp} for more details on these divergences),
a cut at the parton level is necessary in order to avoid them. Physical
observables should be independent of this cut, but they are not.
Solutions to this problem are known and will be briefly described
in sect.~\ref{sect:MEwPS}. However, it must be stressed that even if
the problem is ignored, the combination of tree-level matrix element
generators and showering programs is essential for 1) the optimisation 
of detector designs and 2) analyses based on multi-jet 
configurations (such as SUSY
signals) where the standard showering codes are basically unable to
describe the kinematics of those processes correctly. Recently
this interfacing task has been standardised for FORTRAN-based event
generators by the Les Houches Accord (LHA) event record~\cite{Boos:2001cv}
(the LHA standard is supported in C++ by the HepMC~\cite{Dobbs:2001ck}
event record). The major showering and hadronization programs support 
the LHA and most of the matrix element codes have begun using it.

Tree-level matrix element generators can be divided into two broad
classes, which will be presented in the two subsections below.

\subsection{Matrix Element Generators for Specific Processes}

These codes feature a pre-defined list of partonic processes. The
matrix elements relevant to these processes are obtained with a 
matrix element generation program, which is either part of the package 
or is one of those described in the next subsection. Multi-leg amplitudes 
are strongly and irregularly peaked; for this reason the phase-space 
sampling has typically been optimised for the specific process. The
presence of phase space routines implies that these codes are always
able to output partonic events (weighted or unweighted).

\programAbstract{AcerMC}{B.~Kersevan}
\programInfo{Borut Paul Kersevan and Elzbieta Richter-Was}%
{\cite{Kersevan:2002dd}}%
{http://cern.ch/Borut.Kersevan/AcerMC.Welcome.html}
\programVersion{AcerMC 1.4}

The AcerMC Monte Carlo Event Generator~\cite{Kersevan:2002dd} is
dedicated to the generation of Standard Model background processes at
$pp$ LHC collisions.  The program itself provides a library of the
massive matrix elements and phase space modules for generation of a
set of selected processes: $gg, q \bar q \to t \bar t b \bar b$, $ q
\bar q W(\to l \nu) b \bar b$, $q \bar q W(\to l \nu) t \bar t$, $gg,
q \bar q \to Z/\gamma^*(\to l l) b \bar b$, $gg, q \bar q \to
Z/\gamma^*(\to l l, \nu \nu, b \bar b) t \bar t$ and complete
electroweak $gg \to (Z/W/\gamma^* \to) b \bar b t \bar t$ process.
The hard process event, generated with one of these modules, can be
completed by the initial and final state radiation, hadronisation and
decays, simulated with either the PYTHIA 6.2 or HERWIG 6.5 Monte Carlo
event generator.  Interfaces to both of these generators, based on the
Les Houches Interface Standard~\cite{Boos:2001cv}, are provided in the
distribution version. An additional interface to the
TAUOLA~\cite{Jadach:1993hs} and PHOTOS~\cite{Barberio:1993qi} programs
are also provided with AcerMC version 1.4 and later.  The leading
order matrix element codes have been derived with the help of the
MADGRAPH~\cite{Stelzer:1994ta} package.  The phase-space generation is
based on the multi-channel self-optimising approach as proposed in
Ref.~\cite{Berends:2000fj} for the NEXTCALIBUR event
generator. Additional smoothing of the phase space was obtained by
using a modified ac-VEGAS adaptive algorithm routine in order to
improve the generation efficiency.  The main aim and advantage of the
AcerMC generator is providing an efficient (and therefore fast)
generation of unweighted events, the typical unweighting efficiency
being on the order of 20\%.

The distribution includes the AcerMC library, PYTHIA 6.2, HERWIG 6.5,
HELAS, TAUOLA and PHOTOS libraries and requires CERNLIB for the FFREAD
and PDFlib804 routines. AcerMC has been tested to compile using g77
compiler on RedHat linux (versions 7.2-9.0).

\programAbstract{AlpGEN}{M.~Mangano}

\programInfo{M.L.~Mangano, M.~Moretti, F.~Piccinini, R.~Pittau
and A.~Polosa}{Alpgen documentation: \cite{Mangano:2002ea}. Formalism for ME
evaluation: \cite{Caravaglios:1995cd} and 
\cite{Caravaglios:1998yr}. }{http://mlm.home.cern.ch/mlm/alpgen/}
\programVersion{V1.3.3 (Feb 17, 2004)}

Alpgen is designed for the generation of Standard Model processes in hadronic
collisions, with emphasis on final states with large jet
multiplicities. It is based on the exact leading order evaluation of partonic matrix
elements, with the inclusion of $b$ and $t$ quark masses ($c$ masses
are presented in some cases, where necessary) and $t$ and gauge boson
decays with helicity correlations. 
The code generates events in both a weighted and unweighted
mode, and provides book-keeping utilities for a simple online histogramming of
arbitrary kinematical distributions. Several default spectra, in
addition to a logging of total cross-sections, have been implemented
allowing the user to obtain results after a straighforward compilation
and run. Several routines have been made easily accessible
for the user to define analysis cuts and distributions.
Weighted generation allows for
high-statistics parton-level studies. Unweighted events can be
processed in an independent run through shower evolution and
hadronization programs. Interfaces for processing with HERWIG 6.4 and PYTHIA
6.220 are provided as defaults (interfaces for higher versions can be
provided upon request) using the Les Houches format.

The current available processes are:
\begin{itemize}
	\item $(W\to f\bar{f}') Q\bar{Q}+ N$~jets ($Q$
  being a heavy quark and $f=\ell,q$) with $N\le 4$
  	\item $(Z/\gamma^{*}\to f\bar{f}) \, Q\bar{Q}+N$ ~jets ($f=\ell,\nu$) with
  $N\le 4$ 
	\item $(W\to f\bar{f}') + \mbox{charm} + N$~jets ($f=\ell,q$,
  $N\le 5$) 
	\item $(W\to f\bar{f}') + N$~jets ($f=\ell,q$) and
  $(Z/\gamma^{*}\to f\bar{f})+ N$~jets ($f=\ell,\nu$) with $N\le 6$
	\item $nW+mZ+lH+N$~jets, with $n+m+l+N\le 8$, $N\le3$, including all
  2-fermion decay modes of $W$ and $Z$ bosons, with spin correlations
	\item $Q\bar{Q}+N$~jets, with $t\to b f\bar{f}'$ decays and relative spin
  correlations included where relevant, and $N\le 6$
	\item $Q\bar{Q}Q'\bar{Q}'+N$~jets, with $Q$ and $Q'$ heavy quarks 
(possibly equal) and $N\le 4$
	\item $H Q \bar{Q}+N$~jets, with $t\to b f\bar{f}'$ decays and
  relative spin correlations included where relevant and $N\le 4$
	\item  $N$~jets, with $N\le 6$
	\item  $N\gamma+M$ jets, with $N\ge 1$, $N+M\le 8$
  and $M\le 6$. 
\end{itemize}

The following new classes of processes will appear in V1.4: 
\begin{itemize}
	\item $H+N$~jets ($N\le 4$), with the Higgs produced via the effective $ggH$
vertex
	\item single top production.
\end{itemize}

A suite of up-to-date PDF sets is
available with the code. An interface to LHAPDF will appear soon.
The code is written in F77. A F90 variant of the most CPU-demanding
routines, together with a free F90 compiler suitable for Pentium
architectures, are provided as well. Makefiles with compilation
instructions and datacards for ready-to-use execution are provided.
The code has been validated on several platforms and compilers: 
Linux based PC's,
Digital Alpha Unix, HP series 9000/700, Sun work stations and MAC-OSX
with a {\tt g77} (v2.9 up to 3.4) compiler.
Code and documentation 
updates, as well as detailed bug-fix information and revision
history, are available from the above web page and are distributed via
the Alpgen user mailing list (email to michelangelo.mangano@cern.ch to
join the list). 

%
%
%

\programAbstract{Gr@ppa (GRace At Proton-Proton/Antiproton
  collisions)}{S.~Odaka}
\programInfo{S. Tsuno, S. Shimma, J. Fujimoto,
  T. Ishikawa, Y. Kurihara and S. Odaka.}%
{\cite{Tsuno:2002ce}}%
{http://atlas.kek.jp/physics/nlo-wg/grappa.html}

GR@PPA is a framework to extend the GRACE system to hadron collider
interactions. The extension provides mechanisms to refer to PDFs and to
handle several processes (matrix elements) in a single event
generation run.
The first product based on GR@PPA is GR@PPA\_4b, where all the possible
four b-quark production processes within the Standard Model are
implemented. A new package named GR@PPA\_All includes, in addition,
generators for W+jets from 0 jet up to 3 jets, full six-body top pair,
and so on. These processes are all at the tree level (leading order) and
the generated events are unweighted. See the Web page for further
details and to download the program.

The GR@PPA generators must be interfaced to general-purpose generators
in order to add parton showers and further event evolution. Early
versions were coded so that they could be embedded in PYTHIA 6.1 or
could be used stand-alone, while recent versions such as GR@PPA\_4b 2.01
support the LHA event record~\cite{Boos:2001cv}.  Thus, they can be
embedded in HERWIG 6.5 as well as PYTHIA 6.2 (default). Stand-alone
use is supported as well. The event generation can be controlled in
the same way as the built-in generators if embedded. The default PDF
library is the PYTHIA built-in PDFs in the PYTHIA embedding; PDFLIB or
LHAPDF can be chosen as an option. Either PDFLIB or LHAPDF can be
linked when the HERWIG-embed or stand-alone use is chosen.

The programs are written in Fortran (F77). The PYTHIA or HERWIG
library has to be prepared by users if an embedding option is
chosen. PDFLIB and LHAPDF are not packaged with the program. CERNLIB
is necessary to run sample programs. Each package includes all
the other necessary libraries, a Makefile and instructions for setup
on Unix systems, and sample programs.

\programAbstract{MadCUP}{D. Zeppenfeld}
\programInfo{K~.Cranmer, T.~Figy, W.~Quayle, D.~Rainwater, and
D.~Zeppenfeld.}%
{Apart from the webpage there is as yet no written or published document giving
specifics of the software. Users should cite the web-page~\cite{Madcup}.}%
{http://pheno.physics.wisc.edu/Software/MadCUP/}.

The Madison Collection of User Processes (MadCUP) is a collection of
parton level Monte Carlo programs which have in the past been used for
a variety of phenomenology research papers. The web-page provides
links for downloading the source code and pregenerated event files,
which can be read directly into PYTHIA.  At present, the site provides
Fortran77 source code for production of $W+n$~jets and $Z+n$~jets at
order $\alpha\alpha_s^n$ ($n=2,3$) (dubbed QCW W+2 jet production
etc.) and $W+2$~jet production at order $\alpha^3$ (dubbed EW Wjj
production), i.e. all codes are at LO.  Also available is tree level
code for $t\bar t$ and $t\bar tj$ production.  All codes include
leptonic decay processes of the top quarks and $W,Z$ with full spin
correlations. The $W,Z$ production codes include full off-shell
effects.

All codes fill the common blocks of the Les Houches Interface
Standard~\cite{Boos:2001cv} and thus provide full color and flavor 
information. Parton distributions are obtained by linking to PDFLIB
and histograms are generated with hbook. Minor editing of the source code
is required to change these defaults.

The web-page provides links for downloading the source code and
pre-generated event files, which can be read directly into PYTHIA.

Most of the codes generate a data file of unweighted events which can be 
read as external PYTHIA processes with the MadCUP\_reader (provided as 
source code). A few examples of event data files are provided, see the 
web page for further details.

\programAbstract{Vecbos - W/Z + n jets}{W.~Giele}
\programInfo{ F.A. Berends, H. Kuijf, B. Tausk and W.T. Giele}
{\cite{Berends:1990ax}}%
{http://theory.fnal.gov/people/giele/vecbos.html}

%

VECBOS is a leading order Monte Carlo program for inclusive production
of a W-boson plus up to 4 jets or a Z-boson plus up to 3 jets in
Hadron Colliders. The correlations of the vector boson decay fermions
with the rest of the event are built in. Various parton density
functions are available and distributions can be built in numerically.

The program uses analytic formulae for all tree level amplitudes. These
amplitudes were calculated using recursive techniques developed in
refs.~\cite{Berends:1987me,Berends:1988yn,Giele:1989vp}.

\subsection{Matrix Element Generators for Arbitrary Processes}

The programs described in this subsection may be thought of as automated
matrix element generator authors. The user inputs the initial and final
state particles for a process.  Then the program enumerates Feynman
diagrams contributing to that process and writes the code to evaluate
the matrix element in a programming language such as C or FORTRAN.

The programs are able to write matrix elements for {\it any} tree
level SM process. The limiting factor for the complexity of the events is
simply the power of the computer running the program.  Typically
Standard Model particles and couplings, and some common extensions are
known to the program.

Many of the programs include phase space sampling routines. As such, they
are able to generate not only the matrix elements, but to use those
matrix elements to generate partonic events (some programs also include
acceptance-rejection routines to unweight these events).

\programAbstract{AMEGIC++}{F.~Krauss}
\programInfo{Tanju Gleisberg, Frank Krauss, Ralf Kuhn,
Andreas Sch{\"a}licke, Steffen Schumann, Jan Winter}%
{\cite{Krauss:2001iv} is the {\tt AMEGIC++} manual for version 1.0 
(manual for the improved version 2.0 is in progress).}%
{}
\programVersion{{\tt AMEGIC++} 2.0}

{\tt AMEGIC++} ({\tt A M}atrix {\tt E}lement {\tt G}enerator {\tt I}n
{\tt C++}) is a matrix element generator written in {\tt C++}. It
constitutes an integral part of the new event generator {\tt SHERPA}
({\tt S}imulation for {\tt H}igh {\tt E}nergy {\tt R}eactions of
{\tt PA}rticles) by providing hard tree-level matrix elements and
suitable integrators for $1\to n$ particle
decays and $2\to n$ particle scatterings in the Standard Model,
its minimal supersymmetric extension and an ADD model of extra
dimensions. To evaluate such processes, suitable Feynman diagrams
are generated by {\tt AMEGIC++} and translated into helicity
amplitudes which are then simplified and stored as library files.
The integration over the multi-dimensional phase space is performed
through a multi-channel method with self-adapting weights; the
individual channels are also constructed internally. This is done by
inspection of the Feynman diagrams and mapping out their kinematical
structure in terms of pre-defined building blocks. Finally, the
channels are also written out as library files. Hence it is
appropriate to call {\tt AMEGIC++} a generator-generator, since it
produces complete matrix element generators to run with the core
program. In a first initialization run, these files and the
corresponding makefiles are generated, after compiling and linking a
second run will start the evaluation of cross sections. Due to its
object-oriented structure it is very easy to include new physics
models as long a no new spin states for particles are involved
beyond what is supported at the moment (Spin-0, 1/2, 1, and 2).
Of course, {\tt AMEGIC++} is able to produce weighted and unweighted
events to allow for usage in the framework of event generators.
                                                                               
In the {\tt SHERPA}-framework {\tt AMEGIC++} is interfaced to a full
wealth of other codes, these include:
\begin{itemize}
\item Spectrum generators: Hdecay (for SM Higgs width and branching
  ratios), Isajet/Isasusy (for the MSSM)
\item Laser-Backscattering beam spectrum; own {\tt C++} version of
  the CompAZ parametrization
\item PDF's: LHAPDF, MRST99 ({\tt C++}-version), CTEQ6 (Fortran
  version outside LHAPDF)
\item Parton showers: {\tt APACIC++} with proper ME+PS merging
\item Hadronization and hadron decays: {\tt Pythia 6.163}
\item Event records: HepEvt, HepMC
\end{itemize}

Reference~ \cite{Schalicke:2002ck} describes the implementation of the
  YFS scheme for initial state radiation in lepton
  collisions. Ref.~\cite{Gleisberg:2003ue} provides details on the
  implementation of the ADD model.

\programAbstract{CompHEP}{E.~Boos and S.~Ilyin}
\programInfo{
{\small [Authors for CompHEP 4.2 and later versions]:}
E.~Boos, V.~Bunichev, M.~Dubinin, L.~Dudko, V.~Edneral, V.~Ilyin, A.~Kryukov,
V.~Savrin, A.~Semenov, A.~Sherstnev (the CompHEP collaboration) \\
{\small [Authors for CompHEP 4.1 and earlier versions]:}
A. Pukhov, E. Boos, M. Dubinin, V. Edneral, V. Ilyin, 
  D. Kovalenko, A. Kryukov, V. Savrin, S. Shichanin, A. Semenov}%
{The documentation for CompHEP 4.2 and later versions is not fully ready yet.
Refer to the website and to the ``User's manual 
for version 3.3'',  \cite{Pukhov:1999gg}}%
{http://theory.sinp.msu.ru/comphep}
\programVersion{CompHEP 4.4.0}

CompHEP is a package for evaluating Feynman diagrams, integrating over 
multi-particle phase space and generating events with a high level 
of automation. 

CompHEP allows users to generate Feynman diagrams and to present them
in a graphical form with a Latex output. CompHEP computes squared
Feynman diagrams symbolically and then numerically calculates cross
sections and distributions.  After numerical computation one can
generate with CompHEP the unweighted events with implemented colour
flow information. The events are in the form of the Les
Houches Accord event record to be used in the PYTHIA program for
showering and hadronization with the help of the new CompHEP-PYTHIA
interface. An interface to HERWIG will be available soon. 
CompHEP has an option to introduce new physical models using a friendly
graphical interface to enter new particles and/or new interaction
vertexes or to modify the existing ones. CompHEP 4.4.0 
includes the specialized
package LanHEP \cite{Semenov:2002jw} which allows 
automatically generated Feynman rules 
(the list of propagators and vertexes) for new 
physics models in a standard CompHEP format. CompHEP 4.4.0
includes as the built-in models QED, Fermi model, SM in the unitary
and t'Hooft-Feynman gauges, the variants of the SM models SM\_ud and
SM\_qQ with simplification of light quark combinatorics~\cite{Boos:2000ap}, the
unconstrained MSSM in the unitary and t'Hooft-Feynman gauges, mSUGRA
and GMSB in the unitary gauge with the interface to ISASUSY, and
FeynHiggsFast.  Several other models, like Leptoquark, complete THDM,
Excited Lepton etc. are availble by request.

CompHEP is written in C. It allows for the computation of scattering
processes with up to 6 particles and decay processes with up to 7 particles in
the final state.  However, in practice a computation of a complete set
of diagrams with 6 and 7 final particles takes a lot of time and
computer resources. In this case CompHEP could be used, for instance,
to compute signal contributions taking into account final widths and
spin correlations. Some caution related to the gauge invariance is
needed here.  CompHEP is a the tree level program, so it basically
does computations at leading order. However it allows the inclusion of
partial (approximate) NLO corrections: NLO tree level $2\rightarrow
N+1$ real emission corrections to the $2\rightarrow N$ process
(for example, in high $p_t$ regions if important), NLO structure functions,
loop relations between parameters, known K-factors, and the known loop
contributions as effective vertices. The latter can be done
numerically but not in a fully automatic way.

CompHEP with the interface to PYTHIA, and with the new scripts for symbolic 
and numerical batch modes is a powerful tool for the simulation of different 
physical processes at hadron and lepton colliders. New batch modes provide 
possibilities to use large computer clusters and/or MC farms in a 
parallel way.

One should also stress that the symbolic CompHEP answers for squared
diagrams with the output in the form REDUCE, MATHEMATICA or FORM codes
are available and give a useful theoretical tool for symbolic
manipulations, especially in the case of new models and new
Lagrangians.

\programAbstract{Grace and Grace/SUSY}{Y.~Kurihara}
\programInfo{J. Fujimoto, T. Ishikawa, M. Jimbo, T. Kaneko, K. Kato, 
S. Kawabata, K. Kon, M. Kuroda, Y. Kurihara, Y. Shimizu, H. Tanaka.}%
{\cite{Fujimoto:2002sj}}%
{http://minami-home.kek.jp/}

GRACE/SUSY is a package for generating tree-level amplitudes and
evaluating the corresponding cross sections of processes of the
Minimal Supersymmetric extension of the Standard Model (MSSM). The
Higgs potential adopted in the system, however, is assumed to have a
more general form indicated by the two-Higgs-doublet model. This
system is an extension of GRACE for the Standard Model (SM) of the
electroweak and strong interactions.  For a given MSSM process the
Feynman graphs and amplitudes at tree-level are automatically
created. Integration of the Monte-Carlo phase space by means of the 
BASES~\cite{Kawabata:1995th} 
algorithm gives the total and differential cross sections. When combined
with the SPRING~\cite{Kawabata:1995th} event generator, 
the program package provides us with
the simulation of SUSY particle productions.

\programAbstract{MadEvent and MadGraph}{F.~Maltoni}
\programInfo{MadEvent: F.~Maltoni, T.~Stelzer; 
MadGraph: T.~Stelzer and W.~F.~Long}%
{MadEvent: \cite{Maltoni:2002qb}, MadGraph: \cite{Stelzer:1994ta}}%
{http://madgraph.physics.uiuc.edu}

{\tt MadEvent}~\cite{Maltoni:2002qb} is a multi-purpose, tree-level 
event generator which is powered by the matrix element generator {\tt
MadGraph}~\cite{Stelzer:1994ta}.  In the present version, a
process-dependent, self-consistent code for a specific SM process (at
any collider,{\it e.g.}, $e^-e^+$, $ep$, $pp$, $p\bar p$) is generated
upon the user's request on a web form at {\tt
http://madgraph.physics.uiuc.edu}.  Given a user process, {\tt
MadGraph} automatically generates the amplitudes for all the relevant
subprocesses and produces the mappings for the integration over the
phase space.  This process-dependent information is packaged into {\tt
MadEvent}, and a stand-alone code is produced that can be downloaded
from the web site and allows the user to calculate cross sections and
to obtain unweighted events automatically.  Once the events have been
generated -- event information, ({\it e.g.} particle id's, momenta,
spin, color connections) is stored in the ``Les Houches''
format~\cite{Boos:2001cv}. Events may be passed directly to a shower
Monte Carlo program (interfaces are available for {\tt HERWIG} and
{\tt PYTHIA}) or may be used as an input for combined
matrix-element/shower calculations, such as the one proposed in
Ref.~\cite{Catani:2001cc}.

The code is written in Fortran 77 and has been developed using the 
{\tt g77} compiler under Linux. 
The code is parallel in nature and it is optimized to run on a PC farm. 
At present, the supported batch system is {\tt PBS}. 
The stand-alone codes do not need any external library. 
{\tt LHAPDF} is supported as an option.

Limitations of the code are related to the maximum number of final state
QCD particles.  Currently, the package is limited to ten thousand diagrams 
per subprocess. So, for example, $W$+5 jets which has been calculated, 
is close to its practical limit. At present, only the Standard Model Feynman
rules are implemented and the user has to provide his/her own rules for beyond
Standard Model physics, such as MSSM.

Further information, including examples, a set of benchmark 
cross-sections for hadron colliders, a list of frequently 
asked questions, downloads and updates can be found 
at\\ {\tt http://madgraph.physics.uiuc.edu}.

\section[Higher Order Corrections -- Perturbative QCD Computations]
{HIGHER ORDER CORRECTIONS -- PERTURBATIVE QCD COMPUTATIONS~\protect
\footnote{Contributed by: S.~Frixione.}}\label{sect:NLOcomp}

In this section we shall briefly describe the problems that arise 
when both real- and virtual-emission diagrams are considered in the
context of a perturbative computation. The former class of diagrams
is the one upon which the tree-level matrix element generators of
the previous section are based. Unfortunately, the techniques which
allow a high degree of automatization in the construction of these
codes are not readily extended to the case of virtual diagrams (although
progress is being made on this point). Furthermore, even with analytical
methods, the computation of multi-leg, one-loop amplitudes is a very
difficult problem which is not limited, as in the case of real diagrams,
by CPU power.  Clearly, the situation worsens at two and three loops,
where only a handful of results are presently available. Each N$^k$LO 
computation (where, roughly speaking, $k$ is the number of loops) basically
involves a laborious and ad-hoc procedure. 

Higher-order QCD computations are a highly technical matter.
However, the beginner should feel uneasy not because of technicalities,
but for more fundamental reasons. In fact, the most natural question
is: in a world where hadrons interact producing other hadrons, why 
do QCD theorists spend most of their time talking about and computing
reactions with quarks and gluons?

Let us defer the answer to this question. In fact, let us defer the
treatment of the case of QCD and instead start by explaining how to organise
a next-to-leading order (NLO) computation in the context of an unphysical 
model, whose only virtue is its simplicity. In this one-dimensional 
model, a system (whose nature is irrelevant) can radiate massless
particles (which we call photons), whose energy we denote by $x$,
with $0\le x \le x_s\le 1$, where $x_s$ is the energy of the system
before the radiation. After the radiation, the energy of the system is
$x_s^\prime=x_s-x$. 

In a perturbative computation, the Born term corresponds to no
emissions. The first non-trivial order in perturbation theory
gets a contribution from those diagrams with one and only one
emission, being either a virtual or real photon. These diagrams are
\begin{figure}[h]
\begin{center}
\includegraphics[width=\textwidth,clip]{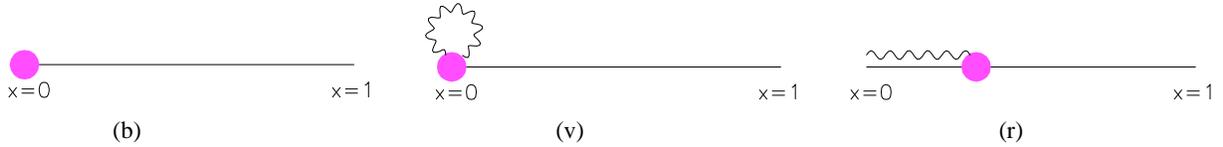}
\end{center}
\caption{\label{fig:toydiag} 
Born (b), virtual (v), and real (r) diagrams for the toy model. The blob
represents the system, the wiggly line the emitted photon.
}
\end{figure}
depicted in fig.~\ref{fig:toydiag}. We write the corresponding contributions 
to the cross section as follows:
\beqn
\xsborn&=&B\delta(x),
\label{born}
\\
\xsvirt&=&a\left(\frac{B}{2\ep}+V\right)\delta(x),
\label{virt}
\\
\xsreal&=&a\frac{R(x)}{x},
\label{real}
\eeqn
for the Born, virtual, and real contributions respectively, where $a$ is
the coupling constant, $B$ and $V$ are constant with respect to $x$, and
\beqn
\lim_{x\to 0}R(x)=B.
\label{limreal}
\eeqn
The constant $B$ appears in eqs.~(\ref{virt}) and~(\ref{limreal}) 
since we expect the residue of the leading singularity of the virtual
and real contributions to be given by the Born term times a suitable
kernel.  (We are cheating a bit here, since we didn't write the Lagrangian
of the toy model from which this property should be derived. We assume it,
since it holds in QCD). We take this kernel equal to 1, since this simplifies 
the computations and it is not restrictive. Finally, $\ep$ is the 
parameter entering dimensional regularization in $4-2\ep$ dimensions.

The task of predicting an observable $O$ to NLO accuracy amounts to 
computing the following integral
\beqn
\VEV{O}=\lim_{\ep\to 0}\int_0^1 dx\,x^{-2\ep} O(x)\left[\xsborn + \xsvirt 
+\xsreal\right],
\label{nlopred}
\eeqn
where $O(x)$ is the observable as a function of $x$, possibly times a 
set of $\stepf$ functions defining a histogram bin. The condition that
the integral of eq.~(\ref{nlopred}) exists is equivalent to the requirement 
that
\beqn
\lim_{x\to 0}O(x)=O(0).
\label{toyIRsafe}
\eeqn
The analogue of this condition in QCD is known as {\em infrared safety}.
The main technical problem in eq.~(\ref{nlopred}) is due to the
presence of the regularising parameter $\ep$.  In order to have an
efficient numerical procedure, it is mandatory to extract the pole in
$\ep$ from the real contribution, thus cancelling analytically
the pole explicitly present in the virtual contribution. One has to keep
in mind that the integral in eq.~(\ref{nlopred}) cannot be fully computed
analytically, because of the complicated form of $O(x)$ and $R(x)$.

Two strategies can be devised to solve this problem.
In the {\em slicing method}, a small parameter $\delta$ is introduced
into the real contribution (third term on the r.h.s. of eq.~(\ref{nlopred}))
in the following way:
\beqn
\VEV{O}_{\sss R}=\int_0^\delta dx\, x^{-2\ep} O(x)\xsreal + 
\int_\delta^1 dx\, x^{-2\ep} O(x)\xsreal .
\label{Rslicing}
\eeqn
In the first term on the r.h.s.\ of this equation we expand $O(x)$ and $R(x)$
in Taylor series around 0 and keep only the first term; the smaller
$\delta$, the better the approximation. On the other hand, the second
term in eq.~(\ref{Rslicing}) does not contain any singularity and
we can just set $\ep=0$ there. We obtain
\beqn
\VEV{O}_{\sss R}&=&aBO(0)\int_0^\delta dx\, \frac{x^{-2\ep}}{x} +
\int_\delta^1 dx\, O(x)\xsreal + {\cal O}(\delta)
\\
&=&a\left(-\frac{1}{2\ep}+\log\delta\right)BO(0)
+a\int_\delta^1 dx\, \frac{O(x)R(x)}{x} + {\cal O}(\delta,\ep).
\eeqn
Using this result in eq.~(\ref{nlopred}), we get the NLO prediction
for $\VEV{O}$ as given in the slicing method:
\beqn
\VEV{O}_{\sss slice}=BO(0)+a\left[\left(B\log\delta +V\right)O(0) + 
\int_\delta^1 dx\, \frac{O(x)R(x)}{x}\right]
+ {\cal O}(\delta).
\label{nloslicing}
\eeqn
The terms collectively denoted by ${\cal O}(\delta)$, although computable, 
can be neglected by choosing $\delta$ small. In practical computations, the 
integral is performed numerically, but due to the divergence of the integrand 
for $x\to 0$, $\delta$ cannot be taken too small because of the loss of
accuracy of the numerical integration. Thus, the value of $\delta$ is
a compromise between these two opposite requirements, being neither
too small nor too large. Of course, ``small'' and ``large'' are
meaningful only when referred to a specific computation. Therefore,
when using the slicing method, it is mandatory to check that the
physical results are stable against the variation of the value of
$\delta$, chosen over a suitable range. In principle, this check would
have to be performed for each observable, $O$, computed.  In practice,
only one observable is considered which is generally chosen to be rather
inclusive (such as a total rate).

In the {\em subtraction method}, no approximation is performed.
One writes the real contribution as follows:
\beqn
\VEV{O}_{\sss R}=aBO(0)\int_0^1 dx\, \frac{x^{-2\ep}\stepf(\xc-x)}{x} 
+a\int_0^1 dx\, \frac{O(x)R(x)-BO(0)\stepf(\xc-x)}{x^{1+2\ep}}\;,
\label{subtraction}
\eeqn
where $x_c$ is an arbitrary parameter $0<x_c\le 1$. The second term on 
the r.h.s.\ does not contain singularities and we can set $\ep=0$ there:
\beqn
\VEV{O}_{\sss R}=-aB\frac{\xc^{-2\ep}}{2\ep}O(0)
+a\int_0^1 dx\, \frac{O(x)R(x)-BO(0)\stepf(\xc-x)}{x}\;.
\eeqn
Therefore, the NLO prediction as given in the subtraction method is:
\beqn
\VEV{O}_{\sss sub}=BO(0)+a\left[(B\log\xc + V) O(0) + 
\int_0^1 dx\, \frac{O(x)R(x)-BO(0)\stepf(\xc-x)}{x}\right].
\label{nlosubt}
\eeqn
This equation has to be compared with eq.~(\ref{nloslicing}).  Although 
the two are quite similar, there are two important differences that have
to be stressed. First, the parameter $x_c$ introduced in the subtraction
method does not need to be small.  (Actually, in the original
formulation of the method $x_c$ was not even introduced, which 
corresponds to setting $x_c=1$ here). This is due to the fact that in
the subtraction method no approximation has been performed in the
intermediate steps of the computation. This in turn implies the
second point; there is no need to check that the physical results
are independent of the value of $x_c$, since this is true by
construction. 

We stress that both eqs.~(\ref{nloslicing}) and~(\ref{nlosubt}) are quite
powerful.  The cancellation of the divergent terms, which arise in the
intermediate steps of the computation from loop and phase-space integrals 
for the case of virtual and real contributions respectively, has been achieved
without knowing anything about: 1) the observable $O$, apart from its infrared
safety, and 2) the matrix elements, apart from their leading singular 
behaviour (see eq.~(\ref{limreal})). 

We now turn to the case of QCD. As anticipated, for the time being we
assume the world to be made of quarks and gluons, and we compute
cross sections for their scatterings. As in the toy model, NLO corrections
imply the computation of virtual and real diagrams. According to the toy
model, in order to achieve the cancellation of the 
singularities\footnote{We are again sloppy here and don't consider 
ultraviolet singularities, which we assume to be properly cancelled by
standard renormalization techniques.} it is crucial to single out the
singular terms in the matrix elements. Let us first consider the case
of real emissions. It is not difficult to realise that the only diagrams
which can contribute a singularity in the matrix elements are those in
which an emission occurs on an external leg (strictly speaking, this is
true only in physical gauges). The final-state emission from a quark
(the case of emission from a gluon is completely analogous)
\begin{figure}[h]
\begin{center}
\includegraphics[width=0.3\textwidth,clip]{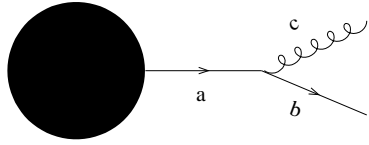}
\end{center}
\caption{\label{fig:Rout} 
Gluon emission from a final-state quark. The blob represents the rest
of the diagram.
}
\end{figure}
can be formally represented as in fig.~\ref{fig:Rout}. The blob represents
the rest of the diagram, which doesn't play any role in what follows and
can be arbitrarily complicated. For the computation of this diagram it is 
convenient to parametrise the momenta as follows:
\beqn
k_b=zk_a+\kt+\zeta_b n\,,\;\;\;\;\;\;
k_c=(1-z)k_a-\kt+\zeta_c n\,,
\label{sudpar}
\eeqn
where $\kt\cdot k_a=\kt\cdot n=0$, $n^2=0$, $k_a^2=0$, $n\cdot k_a\ne 0$, 
and the coefficients $\zeta_b$, $\zeta_c$ are determined by imposing the 
on-shell conditions
\beqn
k_b^2=0\;\Rightarrow\;\zeta_b=-\frac{\kt^2}{2z n\cdot k_a}\,,\;\;\;\;\;\;
k_c^2=0\;\Rightarrow\;\zeta_c=-\frac{\kt^2}{2(1-z) n\cdot k_a}\,.
\eeqn
The computation of the diagram is pretty straightforward and tedious so 
we'll only report the final result.  The contribution to the production cross
section is 
\beqn
d\sigma^{(1,R)}=\frac{\as}{2\pi}\int d\kt^2 dz \,\CF\frac{1+z^2}{1-z}
\frac{1}{\kt^2} d\sigma^{(0)}(k_a) + {\cal R}.
\label{Remout}
\eeqn
with the colour factor $\CF=(N_c^2-1)/(2 N_c) \equiv 4/3$.
A regularization prescription is understood in the first term on the r.h.s. 
of eq.~(\ref{Remout}). The dimensional regularization adopted in the toy 
model would imply an extra factor \mbox{$\kt^{-2\ep}(1-z)^{-2\ep}$}.
Alternatively, the singular regions $\kt\sim 0$, $z\sim 1$ could be
cut off by computing the integral for $\kt>\mu_0$ and $z<1-z_0$.
The quantity $d\sigma^{(0)}(k_a)$ is the cross section computed as if
the outgoing quark $a$ wouldn't have split into $b+c$.  Pictorially,
it corresponds to the square of the black blob, times phase space and
normalization factors. ${\cal R}$ denotes all the terms that are not
singular in $1/\kt$. $d\sigma^{(1,R)}$ is the analogue of
eq.~(\ref{real}) in the toy model, where $x$ plays the role of $\kt$,
and $B$ plays the role of $d\sigma^{(0)}$.  In the toy model, we would
have ${\cal R}=a(R(x)-B)/x$ as a non-singular quantity thanks to
eq.~(\ref{limreal}). Clearly, the structure of eq.~(\ref{Remout})
is more involved than that of eq.~(\ref{real}) since QCD is more
complicated than the toy model. In particular, we see that $d\sigma^{(1,R)}$
is singular not only for $\kt\to 0$ but also for $z\to 1$. 
These two limits correspond to the emitted quark and gluon being collinear,
and to the emitted gluon being soft, respectively. By explicit computation one
thus recovers a well known fact of QCD, that matrix elements are singular when
two on-shell partons become collinear, or a gluon becomes soft.

If QCD works as the toy model, we expect that, upon regulating the 
integral appearing in eq.~(\ref{Remout}), the singularities will cancel 
against those obtained from the virtual contribution. This is in fact what 
happens. The loop integrals can easily be cast in the same form as the 
integral in eq.~(\ref{Remout}):
\beqn
d\sigma^{(1,V)}=-\frac{\as}{2\pi}\int d\kt^2 dz \,\CF\frac{1+z^2}{1-z}
\frac{1}{\kt^2} d\sigma^{(0)}(k_a) + {\cal V}.
\label{Vemout}
\eeqn
This corresponds to writing $\delta(x)/(2\ep)=-\int_0^1 dx x^{-1-2\ep}
+{\cal O}(\ep)$ in the toy model and replacing this expression into 
eq.~(\ref{virt}). The sum of eqs.~(\ref{Remout}) and~(\ref{Vemout}) is 
indeed non singular.

\begin{figure}[h]
\begin{center}
\includegraphics[width=0.3\textwidth,clip]{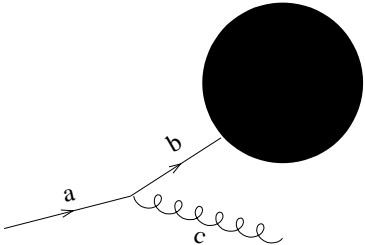}
\end{center}
\caption{\label{fig:Rin} 
Gluon emission from an initial-state quark. The blob represents the rest
of the diagram.
}
\end{figure}

We stress that $d\sigma^{(1,R)}+d\sigma^{(1,V)}$ being
finite is not the analogue of eq.~(\ref{nloslicing}) or eq.~(\ref{nlosubt}).
In fact, here we limited ourselves to computing the most inclusive of
the observables in the kinematics of partons $b$ and $c$, which corresponds
to taking $O\equiv 1$ in the toy model. We shall come back later to the
treatment of non-trivial observables in the case of QCD. Before doing
that, we first have to consider the case in which a splitting occurs
in the initial state. The situation is depicted in fig.~\ref{fig:Rin}.
The kinematics for this splitting can be again parametrized as in
eq.~(\ref{sudpar}). The crucial difference is the momentum entering the 
blob, being in this case $k_a-k_c$. The analogue of eq.~(\ref{Remout}) is
\beqn
d\sigma^{(1,R)}=\frac{\as}{2\pi}\int d\kt^2 dz \,\CF\frac{1+z^2}{1-z}
\frac{1}{\kt^2} d\sigma^{(0)}(zk_a) + {\cal R}.
\label{Remin}
\eeqn
On the other hand, the kinematics of the virtual term are identical to
those relevant to final-state emission (and in fact it is improper to
talk about initial- and final-state contributions to the virtual term).
Therefore
\beqn
d\sigma^{(1,R)}+d\sigma^{(1,V)}=\frac{\as}{2\pi}\int d\kt^2 dz 
\,\CF\frac{1+z^2}{1-z} \frac{1}{\kt^2} 
\left(d\sigma^{(0)}(zk_a)-d\sigma^{(0)}(k_a)\right) + {\cal R} + {\cal V}.
\label{RVemin}
\eeqn
This expression is non-singular in the soft limit $z\to 1$, but it is
still divergent in the collinear limit $\kt\to 0$. We, therefore, have to
conclude that the NLO cross section for a process involving quarks in the 
initial state is a divergent quantity (the same holds true in the case of
gluons in the initial state).

The problem is not as serious as it may seem. To understand this, we 
have to go back to the real world where no one has, or ever ever will,
succeeded in preparing quark or gluon beams. Before QCD was born, Feynman
proposed the {\em parton model} to describe lepton-hadron collisions. 
The cross section for the process $e+H\to X$, $X$ being a generic final
state, is written as
\beqn
d\sigma(K_e,K_H)=\sum_i \int_0^1 dy f^{(H)}_i(y)
d\sigma_i(K_e,yK_H)\,.
\label{eq:pmDIS}
\eeqn
Here, $K_e$ and $K_H$ are the electron and hadron momenta respectively. The 
hadron is thought of as a beam of {\em free massless} 
constituents--the partons--which have only longitudinal (with 
respect to the hadron direction of motion) degrees 
of freedom. There may be different parton types, which are summed over in
eq.~(\ref{eq:pmDIS}). The cross section $\sigma_i$ is relevant to 
the process $e+p_i\to X$, with $p_i$ being a parton of type $i$. Finally, the 
quantity $dy f^{(H)}_i(y)$ is the probability of finding the parton $p_i$ with
3-momentum $\vec{k}_i$ such that $y\vec{K}_H<\vec{k}_i<(y+dy)\vec{K}_H$. The
functions $f^{(H)}_i(y)$ are the parton densities (also called {\em parton
distribution functions}).  They describe a property of the partons as hadron 
constituents and, as such, are independent of the nature of the interactions
between the partons and the lepton.  This property is called {\em
universality}.

In QCD, it appears natural to identify the partons with quarks and
gluons.\footnote{This follows from asymptotic freedom; furthermore, since the
quarks have spin $1/2$ 
and the gluons have zero electric charge, the Callan-Gross relation is
recovered.} The question is: can we compute NLO QCD corrections to
$d\sigma_i$ in eq.~(\ref{eq:pmDIS})? From eq.~(\ref{RVemin}) we
know that the answer is negative, so the parton model does not survive
radiative corrections. However, we shall now show that, by suitably 
modifying it, an equation analogous to eq.~(\ref{eq:pmDIS}) holds, with
all the quantities appearing in it free of divergences. In order to 
proceed, we simplify the notation as follows. We assume that the sum in 
eq.~(\ref{eq:pmDIS}) runs only over quarks so we suppress it and
its dependence upon $i$. We also suppress writing the dependence upon 
$K_e$ and shorten the notation in eq.~(\ref{RVemin}) by introducing the
quantity
\beqn
P(z)=\CF\left(\frac{1+z^2}{1-z}\right)_+.
\eeqn
This notation is denoted as the ``plus prescription'' and is defined 
as follows:
\beqn
\int_0^1 dz h(z)(g(z))_+ = 
\int_0^1 dz (h(z)-h(1)) g(z)\,.
\eeqn
Thus, neglecting the finite terms ${\cal R}$ and ${\cal V}$, we get
\beqn
d\sigma^{(1,R)}(k_a)+d\sigma^{(1,V)}(k_a)=\frac{\as}{2\pi}
\int\frac{d\kt^2}{\kt^2} dz P(z) d\sigma^{(0)}(zk_a) .
\label{RVmod}
\eeqn
We can now compute the $\kt^2$ integral by cutting off the $\kt\sim 0$
region (so we don't use dimensional regularization here):
\beqn
\int\frac{d\kt^2}{\kt^2}\;\longrightarrow\;
\int_{\mu_0^2}^{Q^2}\frac{d\kt^2}{\kt^2}=\log\frac{Q^2}{\mu_0^2}\,,
\eeqn
where $\mu_0$ is an arbitrary mass scale with $\mu_0\ll Q$, and $Q$ is
a characteristic (hard) scale of the process. 
Thus, the NLO contribution to eq.~(\ref{eq:pmDIS}) is
\beqn
d\sigma^{(1)}(K_H)=\frac{\as}{2\pi}\log\frac{Q^2}{\mu_0^2}
\int dy dz f^{(H)}(y) P(z) d\sigma^{(0)}(yzK_H) .
\label{eq:DISNLO}
\eeqn
Since the leading order contribution is
\beqn
d\sigma^{(0)}(K_H)=\int dy f^{(H)}(y) d\sigma^{(0)}(yK_H) ,
\label{eq:DISLO}
\eeqn
after some simple algebra the full NLO prediction can be cast in the 
following form (neglecting terms of order $\as^2$), with
$\mu_0\ll \mu\sim Q$
\beqn
d\sigma(K_H)=\int dy \hat{f}^{(H)}(y,\mu^2,\mu_0^2) 
d\hat{\sigma}(yK_H,\mu^2,Q^2),
\label{eq:DIS}
\eeqn
where
\beqn
\hat{f}^{(H)}(y,\mu^2,\mu_0^2)&=&f^{(H)}(y)+
\frac{\as}{2\pi}\log\frac{\mu^2}{\mu_0^2}
\int_y^1 \frac{dz}{z} P(z) f^{(H)}(y/z)\,,
\label{eq:fhat}
\\*
d\hat{\sigma}(K,\mu^2,Q^2)&=&d\sigma^{(0)}(K)+
\frac{\as}{2\pi}\log\frac{Q^2}{\mu^2}
\int_0^1 dz P(z) d\sigma^{(0)}(zK).
\label{eq:sigmahat}
\eeqn
Eq.~(\ref{eq:DIS}) is meant to replace eq.~(\ref{eq:pmDIS}).  As can be 
seen from eq.~(\ref{eq:sigmahat}), the cross section $d\hat{\sigma}$ is
finite when removing the cutoff, $\mu_0\to 0$, at variance with
eq.~(\ref{RVemin}). However, the cutoff dependence has simply been moved
from the parton cross section to the parton density 
$\hat{f}$~\footnote{In QCD, the notation $f$ is used instead of $\hat{f}$;
here we prefer to use $\hat{f}$ to stress the fact that this quantity is
a subtracted one. Also notice that in QCD $\hat{f}$ is not a probability
density but a number density, since it does not integrate to one.}, which
also acquired in the procedure a dependence on the scale $\mu$.
This is quite an achievement. In fact, the cutoff dependence
is now universal, in the sense that it does not depend upon the nature
of the parton scattering $e+q\to X$.  This means that $\hat{f}$ has the same
universality characteristic of the $f$ originally introduced in the
parton model. Furthermore, the result in eq.~(\ref{eq:sigmahat}) is
likely to be a good approximation of the all-order result, since the
dependence on large scales only implies that the coefficient of 
$\as\equiv\as(\mu)$ is a small number. On the other hand, the same is 
not true in eq.~(\ref{eq:fhat}), since the coefficient of $\as$ is a large
number, $\log\mu^2/\mu_0^2$.  Thus, an all-order computation of $\hat{f}$ would
be necessary in order to obtain a reliable prediction. This is beyond our
current capabilities (although progress is being made in lattice
computations).  However, if we give up the possibility of computing $\hat{f}$,
we may assume we can measure it in a given process and use it to {\em
predict} the cross section of some other process.  This will work because of
the universality of $\hat{f}$.

Eq.~(\ref{eq:sigmahat}) may appear odd because the information on the
production process in the ${\cal O}(\as)$ term is entirely contained
in $d\sigma^{(0)}$, which is of ${\cal O}(\as^0)$. This is because we
have neglected in the derivation the finite terms ${\cal R}$ and ${\cal V}$,
which should be added on the r.h.s. of eq.~(\ref{eq:sigmahat}) to reinstate
the correct notation. It is however important to note that the cancellation
of the divergences proceeds independently of ${\cal R}$ and ${\cal V}$. 
The computation of these finite terms, which is the toughest part of any
matrix element computation, can be forgotten when dealing with the
singularities.

There is possibly a logical flaw in this reasoning, since we didn't
prove that eq.~(\ref{eq:DIS}) holds to all orders. In fact, this is
a highly non-trivial proof, which has been carried out for a 
number of different processes, such as lepton-hadron and hadron-hadron
collisions. The resulting forms, eq.~(\ref{eq:DIS}) or
\beqn
d\sigma(K_{H_1},K_{H_2})=\int dy_1 dy_2 
\hat{f}^{(H_1)}(y_1) \hat{f}^{(H_2)}(y_2)
d\hat{\sigma}(y_1K_{H_1},y_2K_{H_2})
\label{eq:pmDY}
\eeqn
for the collisions of hadrons $H_1$ and $H_2$ with momenta $K_1$ and $K_2$
respectively go under the name of {\em factorization theorems}.

There are a couple of interesting features of the parton densities
$\hat{f}$ that deserve some consideration. First, we consider 
eq.~(\ref{eq:fhat}), by deriving both sides of the equation with
respect to $\log\mu^2$, we get
\beqn
\frac{\partial\hat{f}^{(H)}(y,\mu^2,\mu_0^2)}{\partial\log\mu^2}=
\frac{\as}{2\pi}\int_y^1 \frac{dz}{z} P(z) \hat{f}(y/z,\mu^2,\mu_0^2) 
+{\cal O}(\as^2).
\label{eq:AP}
\eeqn
In this equation, the dependence upon $\mu_0$ is entirely contained in
$\hat{f}$.  Thus, it is sensible to assume the r.h.s. to be the first term
of a well-behaved perturbative expansion (in the sense that the next
terms will be smaller than this). Eq.~(\ref{eq:AP}) is nothing but
the familiar Altarelli-Parisi equation (for the non-singlet case),
often referred to as the Dokshitzer-Gribov-Lipatov-Altarelli-Parisi
(DGLAP) equation.

We can also go back to eq.~(\ref{eq:DISNLO}) and wonder what would have
happened if we had replaced $\mu_0$ with $\mu_1\sim\mu_0$. Clearly
\beqn
\hat{f}^{(H)}(y,\mu^2,\mu_0^2)-\hat{f}^{(H)}(y,\mu^2,\mu_1^2)=
\frac{\as}{2\pi}\log\frac{\mu_1^2}{\mu_0^2}
\int_y^1 \frac{dz}{z} P(z) f^{(H)}(y/z)\,.
\label{eq:fchsch}
\eeqn
The r.h.s. of this equation is a small number. This suggests that the
quantities appearing in eq.~(\ref{eq:DIS}) could have actually been
defined as follows
\beqn
\hat{f}^{(H)}(y,\mu^2,\mu_0^2,\Hfun)&=&f^{(H)}(y)+
\frac{\as}{2\pi}\int_y^1 \frac{dz}{z} 
\left(\log\frac{\mu^2}{\mu_0^2}P(z)+\Hfun(z)\right) f^{(H)}(y/z)\,,
\label{eq:fhatmod}
\\*
d\hat{\sigma}(K,\mu^2,Q^2,\Hfun)&=&d\sigma^{(0)}(K)+
\frac{\as}{2\pi}\int_0^1 dz 
\left(\log\frac{Q^2}{\mu^2} P(z)-\Hfun(z)\right)d\sigma^{(0)}(zK).
\label{eq:sigmahatmod}
\eeqn
The function $\Hfun(z)$ is largely arbitrary, but the idea is that its
contribution to eq.~(\ref{eq:fhatmod}) is a small number compared to
$\log\mu^2/\mu_0^2$. It should be clear that the dependence of $\hat{f}$
upon $\Hfun(z)$ is completely different from that upon $\mu_0$; $\Hfun(z)$ 
is arbitrary, and can be freely changed without changing the l.h.s. of
eq.~(\ref{eq:DIS}). On the other hand, we cannot control the dependence
upon $\mu_0$, which is fixed by Nature.  For this reason, it is customary
to suppress writing it in the arguments of $\hat{f}$.
It should therefore be clear that neither $\hat{f}$ nor $d\hat{\sigma}$
are physical quantities, since they depend on our conventions. Still,
with a given choice of $\Hfun(z)$, $\hat{f}$ is universal and can therefore
be used in the computation of eq.~(\ref{eq:pmDY}), provided that 
$d\hat{\sigma}$ there is defined according to the same conventions
used in eq.~(\ref{eq:sigmahatmod}). The choice of $\Hfun(z)$ is usually
denoted as {\em scheme choice}, with popular schemes, such as $\MSB$ or
DIS, using a specific form for $\Hfun(z)$. We stress again that
not only parton densities, but also parton cross sections are computed 
in a given scheme. An error, of next-to-leading order, is made if one 
predicts an observable using parton densities and parton cross sections 
computed in different schemes.  
Therefore, standard Monte Carlo parton shower codes, which implement hard cross
sections only at the leading order, can be used with parton densities
defined in whatever scheme. On the other hand, care is required with
codes that implement NLO corrections, since the parton density scheme
must match that used in the parton cross sections. 

In summary, with a purely theoretical argument (the presence of collinear
divergences in the cross section for a process involving quarks in the 
initial state) we showed that a world without hadrons cannot exist 
(or at least that QCD is not able to describe it). The 
parton model doesn't survive radiative corrections.  However, its analogue 
in perturbative QCD, the factorization theorem, is based on the very
same physical picture\footnote{Hence, the factorization theorem is
sometimes referred to as QCD-improved parton model, a fairly horrible
name.}. Although hadrons, and not quarks and gluons, are present in
the final state, no final-state collinear divergences are found in
the cross section we previously computed.  This is at variance with
eq.~(\ref{RVemin}), where the sum of eqs.~(\ref{Remout}) and~(\ref{Vemout})
is finite. This prevents us from using quantities akin to parton densities
in the final state, which would convert quarks and gluons into the
observed hadrons (since we have explicitly shown that the introduction of 
parton densities is associated with the presence of collinear divergences).
In order to give a physical meaning to QCD computations, one has to 
{\em assume} that those cross sections which can be computed in terms of 
quarks and gluons, and are free of infrared and collinear divergences,
correspond to cross sections of physical hadrons. This assumption is
know as {\em hadron-parton duality}. For example, the computation at 
${\cal O}(\as^n)$ of the total rate for producing any number of quarks 
and gluons in $\epem$ collisions has to be interpreted as the 
${\cal O}(\as^n)$ prediction for the total $\epem$ hadronic cross section.

It is not difficult to obtain a QCD cross section with final-state
collinear divergences. For example, this would have happened by fixing the 
momentum of the gluon $c$ in fig.~\ref{fig:Rout}. In general, the final-state
quark and gluon momenta are combined in order to define the observable we
want to study. In the previous example, we considered the simplest possible
observable, the total rate, which corresponds to setting $O(x)\equiv 1$ in
the toy model. In general, the observable definition in terms of 
momenta is non trivial. Since the kinematics of the real and virtual 
diagrams are different, so is the definition of the observable.  In the
toy model, the contribution of the real (virtual) diagrams to the observable
$O$ is $O(x)$ ($O(0)$), and the condition of eq.~(\ref{toyIRsafe}) must
be fulfilled in order to obtain a finite cross section. Physically, the
meaning of eq.~(\ref{toyIRsafe}) is clear; the smaller the emitted
photon energy, the closer the value of the observable to the value of
the observable computed when no emissions have occurred. It is easy to
prove that analogous conditions must hold in QCD for avoiding divergences:
the observable value must be insensitive to soft emissions or collinear
splittings. These conditions are known as {\em infrared safety}.

With an infrared-safe jet definition, hadron-parton duality is the understood 
assumption in the extremely successful comparisons between jet data and NLO 
parton-level computations. On the other hand, there are physically 
interesting cases associated with final-state collinear divergent QCD 
cross sections (for example, $\pi^0$ spectra in $p\bar{p}$ collisions).
In these cases, the day is saved by applying parton-model techniques to 
final-state emissions, that is, by proving other factorization theorems. 
The cross section is written analogously to eq.~(\ref{eq:DIS})
\beqn
d\sigma(K_H)=\int dz d\hat{\sigma}_a(K_H/z,\mu^2,Q^2)
\hat{D}^{(a)}_H(z,\mu^2,\mu_0^2),
\label{eq:factfin}
\eeqn
where $K_H$ is now the momentum of the final-state hadron $H$,
$d\hat{\sigma}_a$ is the cross section for producing a final-state
parton $a$, and $\hat{D}^{(a)}_H$ is analogous to $\hat{f}^{(H)}$, which 
is related to the probability density of finding a hadron $H$ within a parton 
$a$, rather than to the probability of finding a parton in a hadron. This
``final-state parton density'' is actually a hadron density and is
called a {\em fragmentation function}. In eq.~(\ref{eq:factfin}) all
the quantities are finite, with those on the r.h.s. being defined through
equations similar to eqs.~(\ref{eq:fhat}) and~(\ref{eq:sigmahat}).
Notice that the finiteness of eq.~(\ref{eq:factfin}) is not in
contradiction with what was said before. In fact, the fragmentation
functions cannot be computed in perturbation theory.  However, similarly
to the parton densities, they can be extracted from data in an universal
manner, i.e. independently of the collision process considered.

It is worth mentioning that the factorization theorems and the hadron-parton
duality hold up to terms proportional to some inverse power of the hard scale
of the process, $1/Q^p$. For this reason, these terms are usually referred to
as {\em power-suppressed terms}. In real experiments, the scale $Q$ may not be
large enough for them to be safely neglected in the comparison with
perturbative QCD predictions.  In the vast majority of cases, they are
estimated with the help of parton shower Monte Carlos, although alternative
approaches exist for $\epem$ and $eH$ collisions.

The factorization theorems and the hadron-parton duality are a rather
powerful machinery, and allow the computation of QCD corrections to
the observable that one wants to compare with data. Nowadays, NLO
predictions exist for the vast majority of observables measured by
experiments and a few NNLO computations are available as well.
Because of the delicate singularity cancellations, early computations
were typically performed for a specific observable in a given collision
process. However, it was later realised that such cancellations basically
rely on properties common to all matrix elements and are independent of
the observable being studied (apart from the requirement that it be
infrared safe). This forms the core of the so-called {\em universal
formalisms} for dealing with infrared singularities.  These formalisms,
at present only available at the NLO, allow one to write any cross
section in terms of finite quantities, obtained with a well-defined
prescription from matrix-element computations (in the example given
before, these finite quantities are $d\sigma^{(0)}$, ${\cal R}$, and 
${\cal V}$). Barring involved technical details, the resulting expressions
are analogous to eqs.~(\ref{nloslicing}) or~(\ref{nlosubt}).   Precisely as in
the toy model, popular universal formalisms rely either on the slicing or on
the subtraction method. They differ in the way in which the slicing parameters
are introduced or in the way in which the subtraction terms are defined.

The universal formalisms can be easily implemented in numerical programs,
which compute $\VEV{O}$ for any observable $O(x)$. From 
eqs.~(\ref{nloslicing}) and~(\ref{nlosubt}), it should be clear that
these programs are simply {\em integrators}, and the integral is typically
computed with Monte Carlo techniques.  The word ``integrator'' is usually
understood or forgotten and ``Monte Carlo'' is what remains.  We shall
refer to them as Monte Carlo Integrators (MCI) in what follows.%
\footnote{These codes have been denoted previously as ``cross section
integrators''. We use the words Monte Carlo here in order to stress the
technique involved in the computation, but no difference of principle is
involved.} The confusion with parton shower Monte Carlos 
(PSMC) is worsened by the fact that one insists that {\em events} are
produced by MCI's, not only by PSMC's. In order to clarify that the
word event is used in MCI's and PSMC's with different meanings, 
consider eq.~(\ref{nlosubt}).  Let's pretend that only the 
last term is present (the remaining ones can be treated similarly) and to
simplify the notation set $\xc=1$. The procedure to obtain 
$\VEV{O}_{\sss sub}$ with Monte Carlo techniques can be summarised as 
follows:
\begin{itemize}
\item Pick at random $0\le x\le 1$.
\item Compute $\wEV=aR(x)/x$ (the {\em event weight}).
\item Compute $\wCNT=-aB/x$ (the {\em counter-event weight}).
\item Call an output routine, that adds $\wEV$ to the bin to which
 $O(x)$ belongs and $\wCNT$ to the bin to which $O(0)$ belongs.
\item Repeat the preceding steps $N$ times and normalise with $1/N$.
\end{itemize}
Picking $x$ at random is equivalent to generating the configuration
$(r)$ in fig.~\ref{fig:toydiag}.  Since this configuration and its 
corresponding weight $\wEV$ are independent of $O(x)$, one calls it
an event.  This is similar to what happens in PSMC, where events are generated
with no reference to an observable. There is more to this similarity,
since the output routine of the fourth step can be called for many
different observables for a given event and eventually predict all 
of them with a single integration procedure. On the other hand,
at variance with PSMC's, each event is accompanied by a counter-event,
whose weight is $\wCNT$ and whose kinematics corresponds to
$(v)$ (or $(b)$, which is equivalent) in fig.~\ref{fig:toydiag}.
Suppose an $x$ value very close to 0 is generated.  The
quantities $\wEV$ and $\wCNT$ will be very large in absolute value and
opposite in sign.  In fact, because of eq.~(\ref{limreal}), they are almost
identical, up to the sign. Thus, if $O(x)$ and $O(0)$ fall in the
same bin, the contributions of the event and of the counter-event tend
to cancel and only a small leftover will contribute to that bin. If, on
the other hand, $O(x)$ and $O(0)$ belong to different bins, the cross
section in these bins will be extremely large in absolute value. Notice
that this can happen also if $O$ is infrared safe, i.e. it fulfils the
condition of eq.~(\ref{toyIRsafe}), so it is sufficient to chose a very
small bin size. One typically says that QCD {\em does not have
infinite-resolution power}.
Finally, let's try to use our MCI as an unweighted-event generator.
According to the hit-and-miss technique (see sect.~\ref{s_simulation}), 
preliminarily one needs to estimate
the maximum of the weight distribution. Since $\wEV$ is divergent for
$x\to 0$, the unweighting procedure is simply undefined. This problem
can be circumvented by introducing a small cutoff (such as $\delta$ of
the slicing method), where the maximum weight will then be $\sim aB/\delta$.
However, the smaller $\delta$, the less efficient the unweighting
procedure. In summary, an MCI is an event generator is the sense that:
{\em a)} the word events implies events and counter-events, as defined
above; {\em b)} only weighted events can be generated and the weights
may have positive or negative values, the latter being typically associated
with counter-events; {\em c)} the events consist of a few final-state quarks 
and gluons (i.e., not hadrons).

\section[Parton Distribution Functions]
{PARTON DISTRIBUTION FUNCTIONS~\protect
\footnote{Contributed by: J.~Huston.}}\label{sect:PDF}

As pointed out in sect.~\ref{sect:NLOcomp}, the calculation of any production
cross sections relies upon a knowledge of the distribution of the momentum
fraction $x$ of the partons (quarks and gluons) in the incoming hadrons in the
relevant kinematic range. These parton densities or parton distribution
functions (PDF's) can not be calculated perturbatively but rather are
determined by global fits to data from deep inelastic scattering (DIS),
Drell-Yan (DY), and jet production at current energy ranges. Two major groups,
CTEQ and MRST, provide semi-regular updates to the parton distributions when
new data and/or theoretical developments become available. The newest PDF's,
in most cases, provide the most accurate description of the world's data, and
should be utilised in preference to older PDF sets. The newest sets from the
two groups are CTEQ6.1~\cite{Stump:2003yu} and MRST2002~\cite{Martin:2002aw}.

\begin{center}
{\em Processes Involved in Global Analysis Fits}
\end{center}

Measurements of DIS structure functions ($F_2,F_3$) 
in lepton-hadron scattering and of lepton pair production cross
sections in hadron-hadron collisions provide the main source of information on
quark distributions $f_q^{(p)}(x,Q^2)$ inside protons.\footnote{The function
$f^{(p)}$ coincides with $\hat{f}^{(p)}$ of sect.~\ref{sect:NLOcomp}.}
At leading order, the gluon distribution function $f_g^{(p)}(x,Q^2)$ enters
directly in hadron-hadron scattering processes with jet final states. Modern
global parton distribution fits are carried out to next-to-leading order which
allows $\as(Q^2)$, $f_q^{(p)}(x,Q^2)$ and $f_g^{(p)}(x,Q^2)$ to all mix and
contribute in the theoretical formulae for all processes.\footnote{This means
that the definition of $\as(Q^2)$ used in a cross section integrator or
event generator needs to be consistent with the specific PDF being employed.}
Nevertheless, the broad picture described above still holds to some degree in
global PDF analyses.

The data from DIS, DY and jet processes utilised in PDF fits cover a wide
range in $x$ and $Q$, but need to be extrapolated to cover the range
accessible at LHC.  HERA data (H1+ZEUS) are predominantly at low $x$, while
the fixed target DIS and DY data are at higher $x$. There is considerable
overlap, however, with the degree of overlap increasing with time as the
statistics of the HERA experiments increases. Parton distributions determined
at a given $x$ and $Q^2$ 'feed-down' to lower $x$ values at higher
$Q^2$ values. DGLAP-based NLO pQCD should provide an accurate description of
the data (and of the evolution of the parton distributions) over the entire
kinematic range currently accessible. 
At very low $x$ and $Q$, DGLAP evolution is believed to
be no longer applicable and a BFKL description must be used.\footnote{
  See e.g.\ Ref.~\cite{Ellis:qj} for a discussion of DGLAP and BFKL.}
No clear
evidence of BFKL physics is seen in the current range of data; thus all global
analyses use conventional DGLAP evolution of PDF's.

There is a remarkable consistency between the data in the PDF fits and the NLO
QCD theory fit to them. On the order of 2000 data points or more are used in
modern global PDF analyses and the $\chi^2$/DOF for the fit of theory to data
is on the order of 1.

The accuracy of the extrapolation to higher $Q^2$ depends on the accuracy of
the original measurement, any uncertainty on $\as(Q^2)$ and the accuracy
of the evolution code.  Current programs in use by CTEQ and MRST should be
able to carry out the evolution using NLO DGLAP to an accuracy of a few
percent over the hadron collider kinematic range, except perhaps at very large
$x$ and very small $x$. Evolution programs are also currently available
which use approximate expressions for NNLO Altarelli-Parisi kernels.

\begin{center}
{\em Parameterizations and Schemes}
\end{center}

A global PDF analysis carried out at next-to-leading order needs to be
performed in a specific renormalization and factorization scheme. The
evolution kernels are in a specific scheme, and to maintain consistency any
hard scattering cross section calculations used for the input processes or
utilising the resulting PDF's need to have been implemented in that same
scheme (see sect.~\ref{sect:NLOcomp}). Almost universally, the $\MSbar$ 
scheme is used, but PDF's are also available in the DIS scheme, a
fixed flavour scheme, and several schemes that differ in their specific
treatment of the charm quark mass.

Some global analyses have also been carried out at
NNLO~\cite{Martin:2002dr,Alekhin:2002fv}. However, the NNLO evolution kernels
are still known only approximately and only the DIS cross sections are known
to NNLO. The other cross sections are still treated at NLO.
	
It is also possible to use only leading-order matrix element calculations in
the global fits which results in leading-order parton distribution
functions. Such PDF's are the standard choice when leading order matrix element
calculations (such as Monte Carlo programs like Herwig and Pythia) are used.
The differences between LO and NLO PDF's, though, are formally NLO.  Thus, the
additional error introduced by using a NLO PDF with Herwig, rather than a LO
PDF, should not be significant, in principle, and NLO PDF's can
be used when no LO alternatives are available (see sect.~\ref{sect:NLOcomp}
for a discussion on this point). The differences between NLO and LO 
parton distributions are not that large for many PDF's in many regions of
$x$ and tend to shrink at higher $Q^2$.

All global analyses use a generic form for the parameterization of both the
quark and gluon distributions at some reference value $Q_o$:
\begin{equation}
F(x,Q_o)=A_ox^{A_1}(1-x)^{A_2}P(x;A_3,...)\,.
\label{eq:PDF}
\end{equation}
The reference value $Q_o$ is usually chosen in the range of 1-2 GeV. The
parameter $A_1$ is associated with small-$x$ Regge behaviour, while $A_2$ is
associated with large-$x$ valence counting rules.  
In general, the first two factors are not sufficient to describe either quark
or gluon distributions. The term $P(x; A_3,...)$ is a suitably chosen smooth
function, depending on one or more parameters, that adds more flexibility to
the PDF parameterization. In general, both the number of free parameters and
the functional form can have an influence on the global fit.

The PDF's made available to the world from the global analysis groups can
either be in a form where the $x$ and $Q^2$ dependence is parameterised or the
PDF's for a given $x$ and $Q^2$ range can be interpolated from either a grid
which is provided or can be generated given the starting parameters for the
PDF's (see the discussion on LHAPDF given below). All of these techniques
should provide an accuracy on the output PDF distributions of the order of a
few percent.

\begin{figure}
\begin{center}
\includegraphics[height=10cm,width=8cm]{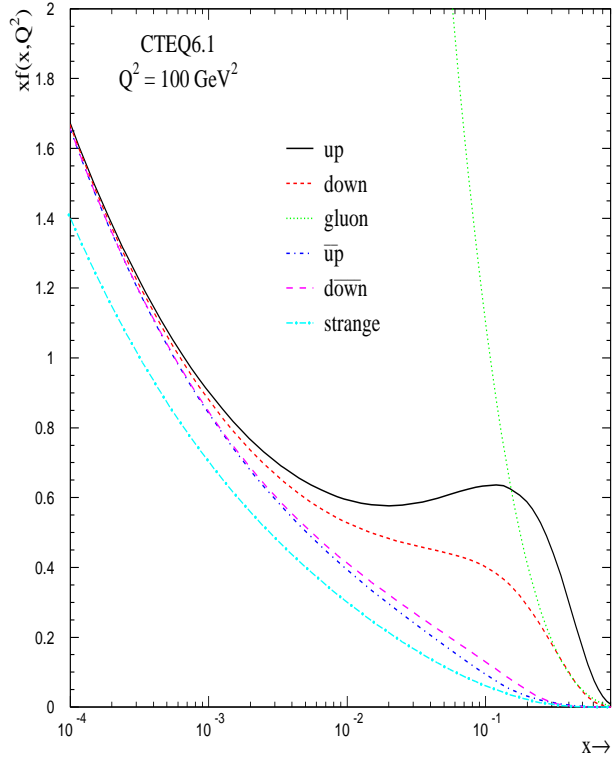}
\end{center}
\caption{
\sf Several PDF's from the CTEQ6.1 set plotted at a $Q^2$ value of 100 $GeV^2$.
} 
\label{fig:AllPdf}
\end{figure}

The parton distributions from the recent CTEQ PDF release are plotted in
Figure~\ref{fig:AllPdf} at a $Q$ value of $100$~GeV. The gluon distribution is
dominant at $x$ values of less than .02 with the valence quark distributions
dominant at higher $x$.

\begin{center}
{\em Uncertainties on PDF's}
\end{center}

In addition to having the best estimates for the values of the PDF's in a
given kinematic range, it is also important to understand the allowed range of
variation of the PDF's, i.e. their uncertainties. A conventional method of
estimating parton distribution uncertainties has been to compare different
published parton distributions.  This is unreliable since most published sets
of parton distributions (for example from CTEQ and MRST) adopt similar
assumptions and the differences between the sets do not fully explore the
uncertainties that actually exist.

The sum of the quark distributions $\left( \Sigma [f_q^{(p)}(x,Q)+
f_{\overline{q}}^{(p)}(x,Q)] \right)$ is, in general, well-determined
over a wide range of $x$ and $Q^2$.  As stated above, the quark distributions
are predominantly determined by the DIS and DY data sets which have large
statistics, and systematic errors in the few percent range ($\pm3\%$ for
$10^{-4}<x<0.75$).  Thus the sum of the quark distributions is basically known
to a similar accuracy. The individual quark flavours, though, may have a
greater uncertainty than the sum. This can be important, for example, in
predicting distributions that depend on specific quark flavours, like the 
$W$ rapidity distribution and its asymmetry.

The largest uncertainty of any parton distribution, however, is that on the
gluon distribution. The gluon distribution can be determined indirectly at low
$x$ by measuring the scaling violations in the quark distributions, but a
direct measurement is necessary at moderate to high $x$. The best direct
information on the gluon distribution at moderate to high $x$ comes from jet
production at the Tevatron.

There has been a great deal of recent activity on the subject of PDF
uncertainties. Two techniques in particular, the Lagrange Multiplier and
Hessian techniques, have been used by CTEQ and MRST to estimate PDF
uncertainties~\cite{Pumplin:2002vw,Martin:2002aw}. The Lagrange Multiplier
technique is useful for probing the PDF uncertainty of a given process, such
as the $W$ cross section, while the Hessian technique provides a more general
framework for estimating the PDF uncertainty for any cross section.

In the Hessian method a large matrix (20x20 for CTEQ, 15x15 for MRST), with
dimensions equal to the number of free parameters in the fit, has to be
diagonalised. The result is 20 (15) orthogonal eigenvector directions for CTEQ
(MRST) which provide the basis for the determination of the PDF error for any
cross section. The larger eigenvalues correspond to directions which are
well-determined.
Each PDF error results from an excursion along the ``+'' and ``-''
directions for each eigenvector. The excursions are symmetric for the
larger eigenvalues, but may be asymmetric for the more poorly
determined directions. There are 40 PDF's for the CTEQ error set and
30 for the MRST error set--one for each eigenvector direction. For a
given event, it is necessary to recalculate the event weight for each
of the error sets in order to evaluate the PDF 
uncertainty.\footnote{This can be a complicated task, as most event
generators are not yet setup to recalculate weights for a given
event with a different PDF set. It is normally not adequate to simply
regenerate a new sample of 
events, as the new events will normally have different kinematics.}

Perhaps the most controversial aspect of PDF uncertainties is the
determination of the $\Delta\chi^2$ excursion from the central fit that is
representative of a reasonable error. CTEQ chooses a $\Delta\chi^2$ value of
100 (corresponding to a 90\% CL limit) while MRST uses a value of 40. Thus, in
general, the PDF uncertainties for any cross section will be larger for the
CTEQ set than for the MRST set. Except at high $x$ ($>0.5$), the uncertainties
on the $u$-quark and $d$-quark distributions are less than 5\%, while the
uncertainty on the gluon distribution is less than 10\% for $x$ values 
smaller than 0.2.

\begin{center}
{\em LHAPDF}
\end{center}

Libraries such as PDFLIB~\cite{Plothow-Besch:1992qj} have been established
that maintain a large collection of available PDF's. However, PDFLIB is no
longer supported making it more difficult for easy access to the most
up-to-date PDF's. In addition, the determination of the PDF uncertainty of any
cross section typically involves the use of a large number of PDF's (on the
order of 30-100) and the manner in which the PDF's are stored in PDFLIB (grids
in $x$ and $Q$) make storage of such ensembles very unwieldy.

At Les Houches in 2001, representatives from a number of PDF groups were
present and an interface was defined (Les Houches accord 2, or 
LHAPDF~\cite{Giele:2002hx}) that allows the compact storage of the information
needed to define a PDF.  Each PDF is determined by only a few lines of
information (basically the starting values of the parameters at $Q=Q_o$) and
the interface carries out the evolution to any $x$ and $Q$ value, at either LO
or NLO as appropriate for each PDF.

The interface is as easy to use as PDFLIB and consists essentially of 3
subroutine calls:

\begin{itemize}
\item call InitPDFset({\it name}): called once at the beginning of the code; 
{\it name} is the file name of the external PDF file that defines the PDF set 
(for example, CTEQ, GKK~\cite{Giele:2001mr} or MRST).
\item call InitPDF({\it mem}): {\it mem} specifies the individual member 
of the PDF set.
\item call evolvePDF({\it x,Q,f}): returns the PDF momentum distributions 
for flavour {\it f} at a momentum fraction {\it x} and scale {\it Q}.
\end{itemize}

The interface can be downloaded at {\tt durpdg.dur.ac.uk/lhapdf/downloads}. 
It is currently included in the matrix element program MCFM (see 
{\tt mcfm.fnal.gov}) and will be included in future versions of the 
cross section integrators and event generator programs. Recent 
modifications make it possible to include all error PDF's in memory at the
same time. Such a possibility reduces the amount of time needed for PDF error
calculations on any observable.

\begin{center}
{\em Resources available}
\end{center}

The PDF's and relevant information can be obtained from the CTEQ and MRST
groups at web addresses given in the references. LHAPDF can be
downloaded from \\
{\tt http://durpdg.dur.ac.uk/lhapdf}. There is also a site where PDF's 
(and their uncertainties) can be displayed on-line:
{\tt http://durpdg.dur.ac.uk/hepdata/pdf3.html}.

\section[Higher Order Corrections -- Showering and Hadronization Event 
Generators]{HIGHER ORDER CORRECTIONS -- SHOWERING AND HADRONIZATION EVENT 
GENERATORS~\protect\footnote{Contributed by: M.~Dobbs.}}\label{s_shg}
\newcommand{\SHG}{SHG}

Programs which employ the parton shower approach, such as PYTHIA,
HERWIG, and ISAJET, have enjoyed widespread use by experimentalists.
These programs, referred to as showering and hadronization generators
(\SHG's), are general purpose tools able to simulate a wide variety of
initial and final states. They begin with a leading order hard
subprocess such as the one ($u\bar{u}\to d\bar{d}$) described in 
sect.~\ref{s_simulation}.
Higher order effects are added by ``evolving'' the event using the parton
shower, which allows partons to split into pairs of other partons
(this splitting is usually denoted as {\em branching} in this context).
The resultant partons are then grouped together or {\it hadronized}
into colour-singlet hadrons and resonances are decayed. Finally, the
underlying structure of the event is generated: beam remnants,
interactions from other partons in the hadrons, and collisions between
other hadrons in the colliding beams (called {\it pile-up}).

\begin{figure}
\centering
\includegraphics[height=.25\textheight,clip=]{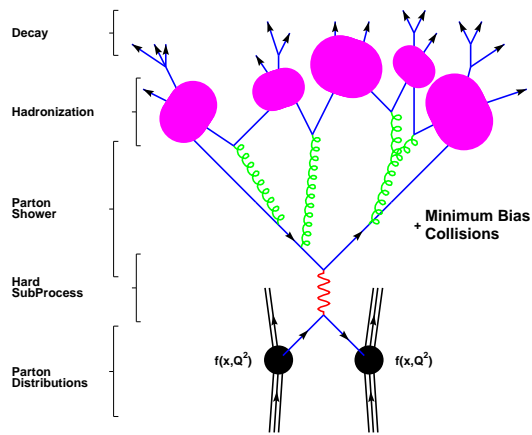}
\caption[Basic structure of a showering and hadronization generator event]
{\label{f_shg_event}
  The basic structure of a showering and hadronization generator event
  is shown schematically\cite{Dobbs:2001ck}.}
\end{figure}

The general structure of the final state of an event from an \SHG\ is shown 
in Figure~\ref{f_shg_event}. The time evolution of the event goes from
bottom to top. Two protons (each indicated by three solid lines to denote
their valence quark content) collide and a parton is resolved at scale
$Q$ and momentum fraction $x$ in each one.  The phenomenology of the
parton resolution is encoded in the parton distribution function
$f(x,Q^2)$. In this example, a valence quark is resolved in the proton
shown on the left, while an anti-quark is resolved from the proton on
the right's sea quark distribution. The quark and anti-quark
annihilate into an $s$-channel resonance denoted by a wavy line. The
resonance then decays into a fermion anti-fermion pair. This part of
the event is called the {\it hard subprocess}. If the resonance is a
$Z^0$ and the initial- and final-state fermion anti-fermion pairs are 
$u\bar{u}$ and $d\bar{d}$ respectively, the physics described in the hard
subprocess is exactly that which is contained in the basic event generator 
of sect.~\ref{s_simulation}. As briefly outlined there, the \SHG\ 
incorporates higher order QCD effects by allowing the (anti)quarks to 
branch into $\qORqbar g$ pairs, while the gluons may branch into $q\bar{q}$ 
or $gg$ pairs. The resultant partons may also branch, resulting in a shower 
or cascade of partons.\footnote{
Though the discussion of parton showers presented here is restricted
to QCD showers, an identical prescription can be applied to
electromagnetic showers and is used in \SHG's to incorporate higher
order QED corrections.}
This part of the event is labelled {\it parton shower} in the figure.
Showering of the initial state partons is also included in the \SHG's,
but is not shown in the figure for simplicity. The event now consists
of a number of elementary particles, including quarks, antiquarks, and
gluons which are not allowed to exist in isolation, as dictated by
colour confinement. Next, the program groups the coloured partons into
colour-singlet composite hadrons using a phenomenological model referred to
as hadronization. The hadronization scale
is in the non-perturbative regime and the programs use fairly crude
phenomenological models, which contain several non-physical parameters
that are tuned using experimental data. Nevertheless, since the
hadronization scale is much smaller than the hard scale(s), the
impact of the hadronization model choice on the final result is
typically small for most physical processes. After hadronization, many
short-lived resonances will be present and are decayed by the
program.

The \SHG's also add in features of the underlying event.  The {\it
beam remnants} are the coloured remains of the proton which are left
behind when the parton which participates in the hard subprocess is
`pulled out'. The motion of the partons inside the proton results in a
small ($\approx1$~GeV) {\it primordial transverse momentum}, against
which the beam remnants recoil.  The beam remnants are colour connected
to the hard subprocess and so should be included in the same
hadronization system.  Multiple parton-parton interactions, wherein
more than one pair of partons from the beam protons interact, are also
accounted for. In a final step, pile-up from other proton-proton
collisions in the same bunch crossing are added to the event.

\SHG's produce events with the frequency predicted by theory, so
they are event generators in the true sense (as opposed to cross section
integrators). One important related point about the generation of an
event with the \SHG's is that, with a few minor exceptions, the hard
subprocess is the only process dependent part. Everything else is
(almost) completely generic and implementing a new physics process
usually only involves implementing the computer code for a new hard
subprocess.\footnote{
  New physical processes can also affect other parts of the
  event, but since we are usually interested in new physics operating at
  large scales, it will have a noticeable impact on the hard
  subprocess only.  }
The \SHG's are normally implemented such that the generation of
everything except the hard subprocess happens with unit
probability---i.e.\ only the hard subprocess has a weight associated
with it. This means (with certain exceptions which are unimportant
here) that after selecting a hard subprocess event using the
hit-and-miss method (see sect.~\ref{s_simulation}), all the other 
aspects of the generation are added
onto the accepted event {\it without ever rejecting the event}.  This
is important for the modularisation of event generators. Thus when an
event generator simulates the hard subprocess a large number of
candidate events are attempted, but only a fraction of those
candidates are accepted. However, for each hard subprocess event that
is chosen and subjected to the subsequent steps of the generation
process, one fully simulated event will come out.

Another important aspect of \SHG's is that they provide an {\it
exclusive} description of the events. As an example, consider the
production of a $Z^0$ boson as the hard subprocess. As already stressed,
at the leading order (i.e., prior to the shower) the transverse momentum of 
the $Z^0$ will always be zero, because there is nothing for the $Z^0$ to 
recoil against. The \SHG's produce transverse momentum for the $Z^0$ through
the parton shower, since the final-state particles emerging from the hard 
subprocess must recoil against those produced by the shower, in order 
to conserve momentum. This prediction of the $Z^0$
transverse momentum is termed {\it exclusive} because of the detailed listing 
(the event record) of the particles recoiling against the $Z^0$ is provided.
In contrast, a cross section integrator results in an {\em inclusive}
prediction because it generally outputs only the $Z^0$ variables and no
information about what the $Z^0$ is recoiling against is provided. Exclusive
calculations---such as those provided by \SHG's---are ideal for the simulation
of experiments, because the full event is necessary for detailed detector
simulation.

\typeout{
The {\it event record} is used by the computer programs to keep track
of the particles and their relationships within the event---and
communicate them to the outside world.  The Fortran
HEPEVT~\cite{HEPEVT} event record is the first standard to be widely
used, and is successful in transferring the information about a fully
generated event to programs for detector simulation. However, it does
not contain enough information to act as an intermediary between two
steps of the generation process.  Recently, this has resulted in
renewed attention for event records.  As the complexity of event
generators grows, it is becoming more common to modularise the
programs, allowing different packages to be responsible for
specialised aspects of the event. A new event record called HEPUP,
developed by Monte Carlo experts (and the author) attending the
2001~Les Houches Physics at TeV Colliders Workshop, has been conceived
and implemented in Ref.~\cite{Boos:2001}. It specifies the protocol
for arbitrary `external' hard subprocesses to be inserted into the
\SHG\ programs. As the high energy physics community moves towards
modularised event generators, it is also moving away from the Fortran
programming language and adopting modern object-oriented languages
like C++. A new C++ event record called HepMC has been co-developed by
the author and J.B.~Hansen~\cite{Dobbs:2001ck}, incorporating and
extending all of the information contained in the Fortran HEPEVT and
HEPUP event records, while employing object-oriented patterns. Both of
these new tools are already being used extensively by the particle
physics community.
}

The most important characteristic for \SHG's is the manner in which
they treat higher order QCD corrections with the parton shower. As
such, this process is described in more detail below. We note that
although some ``predictions'' of the \SHG\ (hadronization, underlying 
event, etc.) have been used in the past in conjunction with NLO cross 
section integrators, these procedures have always been heuristic and
far from being rigorous.  (For example, there is no solid theoretical
argument that justifies the procedure of correcting the NLO parton-level
predictions for jets to the hadron level, which is usually performed
by multiplying the former by the ratio of hadron-level to parton-level 
cross sections in \SHG's).  In particular, the use of the parton shower 
with NLO matrix elements has, until recently, been ``off-limits'' due to
problems with double counting (essentially the corrections will be applied
twice). In sect.~\ref{sect:MEwPS} we discuss a new class of programs which
incorporate NLO matrix elements into SHG's in a consistent manner.

\begin{center}
{\em The Parton Shower~\protect\footnote{Contributed by: B.R.~Webber -- 
for further details and original references see~\cite{Ellis:qj}.}}
\end{center}

\noindent
The parton shower step in Monte Carlo event generation serves two main
purposes:
\begin{itemize}
\item To provide estimates of higher-order corrections that are
enhanced by large kinematic logarithms.  These occur in the phase
space regions of collinear parton branching and/or soft gluon emission;
\item To generate high-multiplicity partonic states which can readily
be converted into the observed hadrons by a soft hadronization mechanism,
i.e.\ one that involves only modest transfers of momentum or quantum
numbers between neighbouring regions of phase space.
\end{itemize}
Schematically, the parton shower is a Markov process\footnote{Basically,
a Markov process is a random process whose future probabilities are 
determined by its most recent values. In other words, if 
$t_1<\ldots <t_n$, we have $P(x(t_n)<x_n|x(t_{n-1}),\ldots,x(t_1))=
P(x(t_n)<x_n|x(t_{n-1}))$.} in which successive
values of an {\em evolution variable} $t$, a {\em momentum fraction} $z$
and an {\em azimuthal angle} $\phi$ are generated, together with the
flavours of the partons emitted during showering.  The evolution
variable $t$ starts at some high value $T$, characteristic of the
hard process, and the next value is selected by solving the equation
\begin{equation}\label{eq:evol}
\Delta_i(T,t_0) = {\cal R}\,\Delta_i(t,t_0)
\end{equation}
where $\Delta_i$ is the {\em Sudakov form factor} for partons of the
relevant flavour $i$, $t_0$ is an infrared cutoff and ${\cal R}\in[0,1]$
is a random number.  The Sudakov form factor is
\begin{equation}\label{eq:delta}
\Delta_i(T,t_0) = \exp\left[-\sum_j\int_{t_0}^T \frac{dt}{t}\int_0^1 dz\,
{\cal P}_{ji}(z,t,t_0)\right]
\end{equation}
where ${\cal P}_{ji}$ is the probability distribution for the parton
branching $i\to j$. Naively, this is given by $\as P_{ji}(z)/2\pi$
where $ P_{ji}$ is the corresponding DGLAP splitting function.
However, in practice the parton branching probabilities are modified
in various ways, the details of which depend on the precise definition
of the evolution variable and the way in which the shower is
implemented:
\begin{enumerate}
\item 
The splitting functions have infrared singularities at $z=0$ and/or 1,
which have to be regularised.  Normally this is done by cutting out the
singular part of the region of integration, in a way that depends on
the evolution variable $t$ and the cutoff $t_0$.  For example, for the
splitting $g\to gg$ in {\small HERWIG} we have
$\sqrt{t_0/t}<z<1-\sqrt{t_0/t}$ (see below).
\item
Quark mass effects may be taken into account in the splittings
$q\to qg$ and $g\to q\bar q$, leading to splitting functions that
depend on quark masses and the evolution variable as well as $z$.
\item
The argument of $\as$ will depend on the evolution scale $t$ and
on the momentum fraction $z$, if important higher-order corrections
are absorbed into the running of the coupling.  The optimal argument,
at least in the case of light partons, is the
relative transverse momentum generated in the splitting.
\item
Other higher-order corrections may be included in the splitting functions.
In this case $1\to 3$ parton splittings can also occur; we ignore
this possibility in the following discussion.
\end{enumerate}
With all these complications, it is impossible to evaluate the integrals
in eq.~(\ref{eq:delta}) in closed form and eq.~(\ref{eq:evol}) cannot be
solved analytically. One can, of course, do numerical integrations and
construct look-up tables of the Sudakov form factors, as is done in
{\small HERWIG}. A neater method, adopted in {\small PYTHIA} and
{\tt Herwig++}, is to make use of the {\em rejection method}.
This involves finding an upper bound
${\cal P}'_{ji} > {\cal P}_{ji}$ for which the integrals can be
done and the equation solved for the next value, $t'$, of the
evolution variable.  Since the Sudakov form factor with ${\cal P}'_{ji}$
in place of ${\cal P}_{ji}$ is a steeper function of $t$, the selected
value $t'$ will tend to be too high.  By accepting this value with
probability ${\cal P}_{ji}/{\cal P}'_{ji}$, and restarting the evolution
with $t'$ in the place of $T$ if it is rejected, one can generate
the correct distribution quite efficiently without any pre-tabulation.

If several types of branching are available for partons of flavour $i$,
for example $g\to q_j\bar q_j$ and $g\to gg$, the next value of
the evolution variable can be selected conveniently by treating
each type separately and selecting the one that chooses the
largest value.  This allows the rejection method to be optimised
separately for each type of branching.

If the selected value of $t$ is less than the cutoff value $t_0$,
i.e.\ if the random number in eq.~(\ref{eq:evol}) is
${\cal R}<\Delta_i(T,t_0)$, then the evolution of parton $i$ has
finished.  It can emit no more (resolvable) partons and is
ready to enter the hadronization stage of the generator.
Depending on the hadronization model, the parton may be set on 
mass-shell or given a virtuality of order $t_0$.

Otherwise,
the next value of the evolution variable $t$ and the type of branching
$i\to j$ having been selected, the momentum fraction $z$ of the
branching is chosen by solving the equation
\begin{equation}\label{eq:zval}
\int_0^z\,dz'{\cal P}_{ji}(z',t,t_0) = {\cal R}'
\int_0^1\,dz'{\cal P}_{ji}(z',t,t_0)
\end{equation}
where  ${\cal R}'\in[0,1]$ is another random number. Here again a
rejection method can be applied, using the upper bound ${\cal P}'_{ji}$
on the branching probability distribution.

Knowledge of $t$ and $z$ at each branching allows (almost) complete
reconstruction of the kinematics of the parton shower. The details
depend on the precise meaning of the shower variables.  In
{\small PYTHIA}, $t$ is the virtuality of the parent parton and
$z$ is a light-cone momentum fraction.  The relative transverse
momentum of the branching (neglecting the virtuality of the
daughters) is then given by $q_t^2 = z(1-z)t$.  On the other hand,
in {\small HERWIG} $t$ represents $E^2(1-\cos\theta)$ where $E$
is the energy of the parent parton and $\theta$ is the opening angle,
while $z$ is an energy fraction, so that $q_t^2 = 2z^2(1-z)^2t$.
In either case, we see that the Sudakov form factor
(\ref{eq:delta}) incorporates the resummation of leading
collinear ($q_t^2\to 0$) singularities to all orders.

The remaining quantity to be fixed at each branching is the
azimuthal angle $\phi$, which fixes the direction of the
relative transverse momentum $q_t$.  This can be chosen with
varying degrees of sophistication. The simplest approach
is to assume a uniform distribution.  More accurately, one
can build in the correct azimuthal correlations between
successive branchings in the collinear approximation.
However, the effect of these is small, since the only
branching with a strong azimuthal correlation is the rare
gluon splitting, $g\to q\bar q$.

Once a branching has occurred, say $i\to jk$ at scale $t_i$, the
evolution of the daughter partons $j$ and $k$ has to be generated.
At the simplest level, their evolution starts at scale $t_i$ and
the next values $t_j$ and $t_k$ are obtained from eq.~(\ref{eq:evol})
using the appropriate Sudakov form factors $\Delta_j$ and
$\Delta_k$, respectively, with $T$ replaced by $t_i$. However,
this implies that $t_j$ and $t_k$ can both be arbitrarily close
to $t_i$, which is impossible.  In {\small PYTHIA}, the
virtualities of the daughters are constrained by the kinematic
relation $\sqrt{t_j}+\sqrt{t_k}<\sqrt{t_i}$. In {\small HERWIG}
the constraint is even stronger, due to {\em angular ordering}.
Recall that in {\small HERWIG} $t_i=E_i^2(1-\cos\theta_i)$, where
$\theta_i$ is the opening angle in the branching $i\to jk$.
Angular ordering means that the opening angle $\theta_j$ of
any subsequent branching of parton $j$ is less than $\theta_i$
and, therefore, $t_j=E_j^2(1-\cos\theta_j) < z^2 t_i$, where
$z = E_j/E_i$.  Hence the evolution of parton $j$ starts at
$z^2 t_i$ rather than $t_i$.  Correspondingly, the evolution
of parton $k$ starts at $(1-z)^2 t_i$.  Note that the condition
for further evolution to be possible is that
$z^2 t_i, (1-z)^2 t_i > t_0$,  which leads to the
condition $\sqrt{t_0/t_i}< z <1-\sqrt{t_0/t_i}$ mentioned above.
In {\small PYTHIA} the angular ordering constraint is
applied subsequently using the rejection method, so its
relation to the shower variables is not so direct.

Angular ordering represents an attempt to simulate more accurately
those higher-order contributions that are enhanced due to soft
gluon emission (and associated virtual corrections).  A soft
gluon emitted by one of the daughter partons in the branching $q\to qg$,
for example, can only resolve the individual outgoing quark and gluon
colour charges if its angle of emission is less than the opening
angle of the branching.  Otherwise, it is emitted by the coherent sum
of their colour charges, which is equal to that of the parent quark.
Therefore we should generate any emission at larger angles from
the parent, not the daughters; this corresponds to angular ordering.
It leads to a suppression of soft gluon emission, which is clearly
reflected in the low-momentum component of hadron jets.

Strictly speaking, whether the daughter colour charges can be resolved
depends on both the azimuthal and the polar angle of emission of the
soft gluon.  Ordering of the polar angles is a valid representation of
soft gluon coherence only after averaging over azimuthal angles.
Therefore it gives results equivalent to resummation of enhanced
soft contributions for observables that are insensitive to
azimuthal distributions, such as the multiplicity distribution 
and single-particle inclusive spectra, but is less precise for
quantities such as the out-of-plane energy flow.

The final outcome of successive branchings is a parton shower in which
each initial parton from the hard process is replaced by a jet of
partons moving in roughly the same direction, together with
some relatively soft wide-angle partons between the jets.  The typical
scale of relative transverse momenta at the end of the shower is set
by the cutoff $t_0$ and not by the scale of the hard process. Furthermore
the shower exhibits {\em preconfinement}: the distribution of colour and
flavour is organised in such a way that non-exotic colour-singlet objects
can form through a soft mechanism involving momentum transfers of
order $t_0$.  Therefore the shower is ideally suited to serve as the
input to a hadronization model.

The approximate treatment of soft gluon coherence by angular ordering
also has implications for the initial conditions of the parton showers.
The maximum angle of emission from a parton $i$ emerging from the hard
process is set by the angle $\theta_{ij}$ between the directions of
that parton and its colour-connected partner $j$, assuming that the
two together form a colour singlet.  Thus, the initial value of the
evolution variable $T$ is in general different for the various partons
involved in the hard process and depends on the colour structure.  In
{\small HERWIG}, for example, if $i$ and $j$ are colour partners we
have $T_i=E_i^2(1-\cos\theta_{ij})$ and $T_j=E_j^2(1-\cos\theta_{ij})$.
These quantities are not separately Lorentz invariant, so the showering
of individual partons is frame-dependent. However, the product
$T_iT_j = (p_i\cdot p_j)^2$ (for massless partons) is invariant,
and the combined shower from the two partons is approximately
frame independent.

Most of the above discussion applies equally well to parton showers
associated with incoming or outgoing legs of the hard process. 
Initial-state showers involve some additional complications due to
the origin of the incoming partons in the colliding beam hadrons.
Evolving downwards from the hard process scale towards the cutoff
corresponds in this case to backward evolution in energy. We then
have to ensure that the energy distribution of the incoming partons
at the cutoff scale is consistent with the measured parton distribution
functions (PDF's) of the incoming hadrons.  This is achieved by weighting
the Sudakov form factors with the PDF's at the corresponding scale.
The different kinematics also mean that the effects of soft gluon
coherence are not so evident in initial-state showers; in fact there
is an enhancement at small momentum fractions rather than a suppression.

\subsection{General Purpose Showering and Hadronization Event Generators}
\label{s_shg_programs}

\programAbstract{HERWIG}{P.~Richardson}
\programInfo{G. Corcella, I.G. Knowles, G. Marchesini, S. Moretti,
  K. Odagiri, P. Richardson, M.H. Seymour, B.R. Webber}%
{\cite{Corcella:2000bw}}{http://hepwww.rl.ac.uk/theory/seymour/herwig/}
\programVersion{6.5}

  HERWIG is a general purpose Monte Carlo event generator for the
  simulation of lepton-lepton, lepton-hadron and hadron-hadron
  collisions. The program includes a large range of hard scattering
  processes together with initial- and final-state radiation using the
  angular-ordered parton shower, hadronization and hadron decays, and
  underlying event simulation.

  The current version of the program, 6.5
  \cite{Corcella:2000bw,Corcella:2002jc}, is available from the HERWIG
  webpage together with the manual, release notes and other
  information. The program includes a Les Houches accord interface to
  allow the user to add new processes and an interface to
  PDFLIB~\cite{Plothow-Besch:1992qj} to allow the use of external
  parton density functions. We also have an interface to
  ISAJET~\cite{Baer:1999sp} for SUSY spectrum and decay rates
  calculations.

%
%
\programsubsection{Subprocesses}

  HERWIG contains a large library of hard $2\to n$ scattering
  processes for both the Standard Model and its supersymmetric
  extension. HERWIG is particularly sophisticated in its treatment of
  the subsequent decay of unstable resonances, including full spin
  correlations for most processes using the approach described
  in~\cite{Richardson:2001df}. 
This method allows us to include simultaneously
  the correct decay matrix element in the decay of these particles,
  the correct correlations both between the production and decay of
  the particles and between all the decays in an event. There is also
  an interface to TAUOLA~\cite{Jadach:1993hs} which allows this
  information to be passed to TAUOLA to include the correct
  polarization in the decay of the taus.

  The following types of process are included:
\begin{description}
\item[QCD] $2\to2$ scattering processes including heavy flavour production,
\item[Electroweak] $\gamma/\gamma^*/Z^0/W^\pm/H^0$ production either singly or in
                   pairs and often with additional hard jets,
\item[SUSY] A large range of MSSM production processes in lepton-lepton and
            hadron-hadron collisions including Higgs production and the option
            of R-parity violating decays and hard production processes,
\item[Exotics] New gauge bosons and resonant graviton production.
\end{description}

  It is unlikely that any additional processes will be added to the Fortran program
  at this point. Any additional processes can now be added by the user using the
  Les Houches Accord.

%
%
\programsubsection{Parton Shower}

  Following the hard scattering process additional QCD radiation is
  generated in HERWIG using a coherent branching algorithm for both
  the initial- and final-state particles. In this algorithm the full
  phase space for emission is restricted to an angular-ordered region
  in order to treat both the leading soft and collinear
  singularities. The simulation also includes azimuthal correlations
  due to spin effects~\cite{Knowles:1988hu,Knowles:1988vs} in the
  parton shower and the dead-cone effect for radiation from massive
  quarks.

  In addition to the parton shower simulation, matrix element
  corrections~\cite{Seymour:1994we,Seymour:1994df} are included for $e^+e^-$
  collisions~\cite{Seymour:1991xa}, deep inelastic scattering
  processes~\cite{Seymour:1994ti}, top quark
  decay~\cite{Corcella:1998rs} and Drell-Yan
  production~\cite{Corcella:1999gs}. This correction consists of two
  parts: the first fills the dead-zone\footnote{This is the region of
  phase space which is not filled by the HERWIG parton shower.}
  according to the leading-order matrix element; while the second
  corrects the emission of any radiation inside the region already
  filled by HERWIG which is capable of being the hardest emission
  according to the leading-order matrix element.
 
%
%
\programsubsection{Underlying Event}

  The underlying event model inside HERWIG is based on the
  minimum-bias pp event generator of the UA5
  Collaboration~\cite{Alner:1986is}, modified to make use of the
  cluster fragmentation algorithm.  In addition to this model there is
  an external package,
  JIMMY~\cite{Butterworth:1993ig,Butterworth:1996zw}, which uses a
  multiple scattering model for the underlying event. Hopefully this
  model will be incorporated into the program in the near future.

%
%
\programsubsection{Hadronization and Hadron Decays}

  HERWIG uses the cluster hadronization model which is based on the
  colour pre-confinement property of the angular-ordered parton
  shower.  After the parton shower phase, any gluons are split
  non-pertubatively into $q\bar{q}$ pairs. In the $N_C\to\infty$ limit,
  all the quarks and antiquarks can be uniquely formed into colour
  singlet clusters; due to colour pre-confinement, the mass spectrum of
  these clusters is strongly peaked at low mass and falls off
  rapidly. The high mass clusters are first split into lower mass
  clusters using a string-like mechanism.  This is followed by the decay of the
  low mass clusters, according to phase space, into the observed
  hadrons.

  The unstable primary hadrons are then decayed. In most cases these
  decays are performed according to phase space with matrix elements
  in only a few special cases.  Interfaces are provided to use
  external packages for the decay of B hadrons.

\programAbstract{Herwig++}{B.R.~Webber}
\programInfo{S.~Gieseke, A.~Ribon, P.~Richardson, M.H.~Seymour,
P.~Stephens, B.R.~Webber}%
{\cite{Gieseke:2003hm}}
{http://www.hep.phy.cam.ac.uk/theory/Herwig++/}
\programVersion{1.0}

Herwig++ is a completely new event generator, written in C++. It is
built on the experience collected with the well-known Fortran event
generator HERWIG, but is not simply a translation. The aim is to
provide a multipurpose event generator with similar or improved
capabilities, such as angular-ordered parton evolution and the cluster
hadronization model, but with greater flexibility, generality and ease
of maintenance. From now on the development of Fortran HERWIG will
cease (apart from bug fixes) and Herwig++ will gradually take over.
 
The main stages of the simulation are the same as in HERWIG.
However, in comparison to its predecessor, Herwig++ features a new parton
shower and an improved cluster hadronization model.  The parton shower
evolution is carried out using new evolution variables suited to describing
radiation from heavy quarks as well as light partons~\cite{Gieseke:2003rz}.
The cluster hadronization model avoids some shortcomings of the
model used in HERWIG and gives yields of baryons and strange
particles in better agreement with LEP data.

A detailed manual for Herwig++ is in preparation~\cite{manual}.  The
program is based on the Toolkit for High Energy Physics Event
Generation 
(ThePEG)~\footnote{\href{http://www.thep.lu.se/ThePEG/}{http://www.thep.lu.se/ThePEG/}}
and the Class Library for High Energy Physics (CLHEP)~\cite{CLHEP}.
They are utilized in order to take advantage of the extended general
functionality they can provide.  The use of ThePEG unifies the event
generation framework with that of
Pythia7~\footnote{\href{http://www.thep.lu.se/Pythia7/}{http://www.thep.lu.se/Pythia7/}}.
This will provide benefits for the user, as the user interface, event
storage etc.\ will appear the same.  The implementations of the
physics models, however, are completely different and independent from
each other.

Version 1.0 of Herwig++ does not contain initial-state parton showering
or a model for the underlying event.  These will be available shortly
in version 2.0. Meanwhile, the program is being tested against
 a wide variety of electron--positron data from LEP and
SLC~\cite{Gieseke:2003hm}.

\programsubsection{Parton shower}

The partonic evolution from the large scale
of the hard collision process down to hadronic scales via the coherent
emission of partons, mainly gluons, is simulated on the basis of the
Sudakov form factor.  Starting from the hard process scale $Q$,
subsequent emissions at scales $Q_i<Q$ and momentum fractions $z_i$
are randomly generated as a Markov
chain on the basis of the soft and collinear approximation to partonic
matrix elements.
In Herwig++ we have chosen a new framework of variables, generically
called $(\tilde q, z)$.  Here, $\tilde q$ is a scale that appears
naturally in the collinear approximation of massive partonic matrix
elements and generalizes the evolution variable of HERWIG to the
evolution of massive quarks. The variable $z$ is a relative momentum
fraction; the evolution is carried out in terms of the Sudakov
decomposition of momenta in the frame where the respective colour
partners are back-to-back.  As in HERWIG, the use of the new variables
allows for an inherent angular ordering of the parton cascade, which
simulates coherence effects in soft gluon emission.  The details of
the underlying formalism are described in ref.~\cite{Gieseke:2003rz}.

The most important parameters of the parton shower are the QCD scale
$\Lambda_{\rm QCD}$ and the cutoff parameter $Q_g$, which
regularizes the soft gluon singularity in the splitting functions and
determines the termination of the parton shower.  Less important but
relevant in extreme cases is the treatment of the strong coupling
constant at low scales.  We have parametrized $\alpha_S(Q)$ below a
small scale $Q_{\rm min} > \Lambda_{\rm QCD}$ in different ways.  We keep
$Q_{\rm min}$ generally to be of the order of 1\,GeV, where we expect
non-perturbative effects to become relevant.  Below that scale
$\alpha_S (Q)$ can optionally be set to zero, frozen, or 
interpolate linearly or quadratically in $Q$, between 0 and
$\alpha_S(Q_{\rm min})$.

\programsubsection{Hadronization and decay} 

We put the final partons of the shower evolution on their constituent
mass shells, since the non-perturbative cluster hadronization will
take over at the cutoff scale.
The partonic final state is turned into a hadronic final state within
the general framework of the cluster hadronization model of
HERWIG~\cite{Webber:1983if}.  In order to address some
shortcomings of the HERWIG model~\cite{kupco},
a new cluster hadronization model has been created for Herwig++.
The method for flavour selection in cluster decays has been
changed so that the probability of choosing a given light hadron
is not reduced when heavier states are added to the particle
tables. In addition, the meson and baryon sectors are treated
separately, and the baryon to meson ratio can be controlled
by the diquark weight parameter. Details can be found in
ref.~\cite{Gieseke:2003hm}.

The emerging hadrons are possibly unstable and eventually decay. At
present the decay matrix elements and modes correspond to those in HERWIG.
A more sophisticated treatment including polarization correlations is
under development for version 2.0.

\programAbstract{ISAJET}{H.~Baer}
\programInfo{F. Paige, S. Protopopescu, H. Baer and X. Tata}
{\cite{Baer:2003mg}}{http://www.phy.bnl.gov/$\sim$isajet}
\programVersion{7.69}

ISAJET is a Monte Carlo program which simulates $pp$, $p\bar{p}$, 
and $e^+e^-$ interactions at high energies\cite{Baer:2003mg}. 
It is based on perturbative QCD plus phenomenological models for 
parton and beam jet fragmentation. 
The manual describes the physics and explains how to use the program.
The code includes a toy calorimeter simulation (CALSIM) and jet 
finder (GETJET).

ISAJET is written in Fortran 77 and is distributed using the 
Patchy code management system developed at CERN. 
The Patchy source file isajet.car can be be unpacked and compiled 
on any supported Unix system by editing the Makefile and 
selecting the appropriate options. 
Compiling ISAJET on any other computer with ANSI Fortran 77 and Patchy, 
including any for which CERNlib is supported, should be straightforward. 
The files isajet.car and Makefile are available via HTTP, 
via anonymous FTP from ftp.phy.bnl.gov/pub/isajet or 
via AFS from /afs/cern.ch/user/p/paige/public/isajet. 
The alternative sources also contain some additional files. 

\programsubsection{Subprocesses}

ISAJET can be used to generate events for all Standard Model
$2\to 2$ subprocesses. Subprocess reactions are controlled by
specifying the reaction type in the \verb|input.par| file,
where program inputs are stored. Reaction types for hadron colliders include:
TWOJET (quark and gluon production), DRELLYAN ($W$ and $Z$ 
production), WPAIR ($W^+W^-$, $ZZ$ $WZ$, $W\gamma$ and $Z\gamma$ production
including spin correlations), HIGGS ($s$-channel Higgs boson
production via $q\bar{q}$, $gg$ or vector boson fusion), WHIGGS ($WH$ or
$ZH$ production), PHOTON ($\gamma q$, $\gamma g$ or $\gamma\gamma$
production), SUSY (all lowest order $2\to 2$ sparticle production processes),
TCOLOR (techni-rho production), EXTRDIM (graviton production
in models with large extra dimensions) and MINBIAS (minimum bias events
generated using an $n$-cut Pomeron model with modified hadronization). 
The reaction ZJJ for $Z$ plus 2-jet production has been included, as well
as a first attempt at including $2\to n$ subprocesses.
If DRELLYAN reactions are invoked, the $W$s or $Z$ can be
created as $2\to 1$ subprocesses, or as $2\to 2$ subprocesses as
$Wg$, $Wq$, $Zg$ and $Zq$ production if non-zero PT is stipulated
in the input file. PDFLIB is included if the appropriate link to 
CERNLIB is made.

For $e^+e^-$ colliders,
all SM $2\to 2$ particle and Higgs production processes are included, 
along with $2\to 2$ SUSY particle and SUSY Higgs production processes.  
The $e^+e^-$ reactions can be run with arbitrary electron or positron
beam polarization. In addition, it is an option to run using
electron and photon PDF's from bremsstrahlung. 
Electron and photon beamstrahlung distributions are included as well.

\programsubsection{Parton shower}

Isajet uses the original Fox-Wolfram parton shower algorithm \cite{fw}
for QCD radiation from final state quarks and gluons. In addition,
radiation of $W$s, $Z$s and $\gamma$s from final state particles is
treated in the same approximation.

Radiation from initial state quarks and gluons is invoked using Sj\"ostrands
backward shower algorithm\cite{Sjostrand:1985xi}, 
which actually uses the PDF's to calculate
emission probabilities.

\programsubsection{Hadronization and decays}

Isajet uses a modified Field-Feynman independent hadronization model\cite{Field:fa}
to convert quarks and gluons into mesons and baryons. 
Independent fragmentation correctly describes the fast hadrons in
a jet, but it fails to conserve energy or flavor exactly. Energy
conservation is imposed after the event is generated by boosting the
hadrons to the appropriate rest frame, rescaling all of the
three-momenta, and recalculating the energies.

Unstable particles are decayed further, with decay modes listed in the 
decay table ISADECAY.DAT. ISAJET keeps track of $\tau$ lepton
helicities, and decays the $\tau$s according to weak interaction
decay matrix elements. Exact decay matrix elements are also invoked for
$t$-quark decays and for $3$-body sparticle decays.

\programsubsection{Underlying event}

There is now experimental evidence that beam jets are different in
minimum bias events and in hard scattering events. ISAJET therefore uses
a similar algorithm but different parameters in the two cases.

The standard models for particle production are based on pulling
pairs of particles out of the vacuum by the QCD confining field,
leading naturally to only short-range rapidity correlations and to
essentially Poisson multiplicity fluctuations. The minimum bias data
exhibit Koba-Nielsen-Olesen (KNO) 
scaling and long-range correlations. A natural explanation
of this was given by the model of Abramovskii, Kanchelli and Gribov\cite{akg}.
In their model the basic amplitude is a single cut Pomeron with
Poisson fluctuations around an average multiplicity $\langle n
\rangle$, but unitarity then produces graphs giving $K$ cut Pomerons
with multiplicity $K\langle n \rangle$.
A simplified version of the AKG model is used in ISAJET. The
number of cut Pomerons is chosen with a distribution adjusted to fit the
data. Each cut Pomeron is hadronized in its own
center of mass using a modified independent fragmentation model with
an energy dependent splitting function to reproduce the rise in
$dN/dy$.

\programsubsection{Supersymmetry}

Supersymmetric scattering events can be generated in a wide variety
of SUSY models in ISAJET. A weak scale MSSM model may be invoked,
which assumes $R$-parity conservation and no $CP$ violating phases or
off-diagonal soft SUSY breaking masses. Sparticle masses are computed, and
all sparticle and Higgs boson cascade decay branching fractions are calculated.
The mass spectra and decay table can be output independently via the
(independent) ISASUSY program.

Alternatively, the program ISASUGRA contains a variety of SUSY models
(mSUGRA, minimal and non-minimal GMSB models, non-universal SUGRA,
AMSB model, right-hand neutrino SUGRA model) which require
an iterative solution to the SUSY renormalization group equations (RGEs).
ISASUGRA includes 2-loop RGEs for both couplings and soft SUSY breaking terms.
Electroweak symmetry is broken radiatively, and the renormalization 
group improved 1-loop effective potential is minimized at the high
scale $Q=\sqrt{m_{{\tilde t}_R} m_{{\tilde t}_L}}$, which accounts for 
leading 2-loop terms in the computation of $\mu$ and the SUSY Higgs boson
masses. All sparticle masses are calculated including full 1-loop 
radiative corrections. Sparticle masses and the decay table are output
by the (independent) ISASUGRA program. 

ISASUSY or ISASUGRA inputs can be included in the \verb|input.par|
file for sparticle or Higgs boson event generation within these
scenarios. $R$-parity violation decays may be simply included
by adding these to the ISADECAY.DAT file with the appropriate 
branching fractions.

\programAbstract{PYTHIA}{T.~Sj\"ostrand and P.~Skands}
\programInfo{T.~Sj\"ostrand, L.~L\"onnblad,
S.~Mrenna and P.Z.~Skands}%
{Please cite the latest published edition, \cite{Sjostrand:2000wi}}%
{http://www.thep.lu.se/$\sim$torbjorn/Pythia.html}
\programVersion{Use stable version 6.222 for production, beta version 6.303.}

\textsc{Pythia} is a general-purpose generator for hadronic events
in pp, e$^+$e$^-$ and ep colliders. It contains a subprocess library
and generation machinery, initial- and final-state parton showers,
underlying event, hadronization and decays, and analysis tools. The
physics aspects are described separately in the subsections below.
 
\textsc{Pythia} was combined with \textsc{Jetset} in 1997, to form a 
single, self-contained library. The current version can be downloaded
from the Pythia webpage, where you can also find the 
manual~\cite{Sjostrand:2001yu}, update notes, sample main programs, an
archive of previous versions, and more.
 
The program is written entirely in Fortran 77; there are plans to move
to C++. Particle codes are given in the PDG standard. Parton-level
configurations can be input from the Les Houches Accord Event Record,
and hadron-level events can be output to (or input from) the HEPEVT
commonblock. PDFLIB is interfaced, and via this interface LHAPDF
can also be used. 
An interface to the Les Houches Accord SUSY spectrum and decay
calculations~\cite{Skands:2003cj} is available in \textsc{Pythia} 6.3.
 
\programsubsection{Subprocesses}

\textsc{Pythia} contains around 240 different $2\to n$ subprocesses,
all at leading order. Most of these are $2 \to 2$, some $2 \to 1$ or 
$2 \to 3$. The subsequent decays of unstable resonances ($W$, $Z$, top, 
Higgs, SUSY, \ldots) brings up the partonic multiplicity, for many 
processes with full spin correlations in the decays. The physics areas 
covered include:\\
-- {QCD:} $2 \to 2$ partonic scattering, heavy flavour, elastic and
diffractive processes;\\
-- {Standard Model:} $\gamma/\gamma^*/Z^0/W^{\pm}$ singly or in pairs,
or with a quark or gluon, Higgs;\\
-- {SUSY:} two Higgs doublets, sfermion and gaugino pairs,
$R$-parity-violating decays;\\
-- {Exotics:} Technicolor, new gauge bosons, compositeness, leptoquarks,
doubly charged Higgses, extra dimensions.\\
These internal processes can be mixed freely with Les Houches Accord
external ones, and are, normally evolved through the showering and
hadronization identically.

\programsubsection{Parton Showers}
 
Given the generation of the basic partonic processes listed above, initial- and
final-state showers are added to provide more realistic multipartonic
configurations, especially for the internal structure of jets.
 
The final-state shower \cite{Bengtsson:1986et,Norrbin:2000uu} 
is based on forward evolution in terms of a
decreasing timelike virtuality $m^2$, with angular ordering imposed
by veto. The framework is leading-log, but includes many NLL aspects
such as energy--momentum conservation, $\alpha_s(p_{\perp}^2)$
and coherence. Further features include gluon polarization effects
and photon emission. While of leading-order character for $2 \to 2$
processes, it is matched to first-order (``NLO'') matrix elements for
gluon emission in $1 \to 2$ resonance decays in the Standard Model
and its Minimal Supersymmetric extension, e.g. $t \to b\,Wg$,
$h^0 \to b\,\overline{b}\,g$ and
$\tilde{g} \to \tilde{q}\,\overline{q}\,g$.
 
The initial-state shower \cite{Sjostrand:1985xi,Miu:1998ju} 
is based on backwards evolution, i.e. starting
at the hard scattering and moving backwards in time to the shower
initiators, in terms of a decreasing spacelike virtuality $Q^2$.
It also includes some coherence effects and uses
$\alpha_s(p_{\perp}^2)$. It has been matched to first-order
matrix elements only for $\gamma^*/Z^0/W^{\pm}$ production
(and to $gg \to h^0$ in the heavy-top limit). Partons radiated in
the initial state may initiate final-state showers of their own.
 
Initial and final showers are matched to each other by maximum
emission cones.
 
\programsubsection{Underlying event}
 
The composite nature of hadrons (and resolved photons) allows for several
partons from each of the incoming hadrons to undergo scatterings.
Such multiple parton--parton interactions are, in the \textsc{Pythia}
framework \cite{Sjostrand:1987su}, instrumental in building up the 
activity in the underlying
event, in everything from charged multiplicity distributions and
long-range correlations to minijets and jet pedestals. The
interactions are described by perturbation theory, approximated
by a set of more or less separate $2 \to 2$ scatterings; energy
conservation and other effects introduce (anti)correlations. The
scatterings are colour-connected with each other and with the
beam remnants.
 
The key parameter is a $p_{\perp\mathrm{min}}$ cutoff of the order
of 2~GeV, below which colour screening in the incoming hadrons
is taken to strongly dampen the naive perturbative interaction rate.
Further parameters are related to an assumed impact-parameter
dependence (central vs. peripheral collisions), the primordial
$k_{\perp}$ and energy sharing when there are several partons in
the beam remnants, and so on.
 
Studies are underway to further improve the realism of this
framework~\cite{Sjostrand:2004pf}..
 
\programsubsection{Hadronization and decays}
 
The Lund string model \cite{Andersson:ia,Sjostrand:1984ic} 
is probably the most successful and widely
used framework to understand the hadronization process. It is based
on a picture with linear confinement, where (anti)quarks or other
colour (anti)triplets are located at the ends of the string, and
gluons are energy and momentum carrying kinks on the string. Thereby
a gluon comes to be attached to two string pieces, one related to
its color and the other to its anticolour, and experiences
a confinement force twice that of a quark, just like in the
$N_C \to \infty$ limit of QCD. The string breaks by the production
of new $q\overline{q}$ pairs, and a quark from one break can combine
with an antiquark from an adjacent one to form a colour singlet
meson. The whole framework is very constrained in terms of its
energy--momentum structure, but the flavour selection involves a
multitude of parameters.
 
Unstable particles are allowed to decay. In cases where better
decay models are available elsewhere, e.g. for $\tau^{\pm}$ with
spin information or for $B$ hadrons, such decays can be delegated to
specialized packages.
 
Further components of the hadronization/decay framework include
junctions where three colour lines meet, the special description of
occasional low-mass strings, Bose--Einstein effects among identical
mesons, and colour reconnection effects.

\programAbstract{SHERPA}{F.~Krauss}
\programInfo{Tanju Gleisberg, Frank Krauss, Andreas Sch{\"a}licke,
Steffen Schumann, Jan Winter}%
{A manual is in preparation.}%
{ }
\programVersion{The code is about to be released in an $\alpha$
  version.}

{\tt SHERPA} ({\tt S}imulation for {\tt H}igh {\tt E}nergy {\tt R}eactions
of {\tt PA}rticles) is a new multi purpose event generator for
the simulation of events at lepton and hadron colliders. To a large
extent it is being developed completely independent of the other two
projects {\tt Pythia7} and {\tt HERWIG++} and of structures like
{\tt CLHEP}. In its current state, {\tt SHERPA} includes:
\begin{itemize}
\item The full width of service methods needed, such as an internal
      event record, particle data, four vectors, I/O handling, etc.;
\item A physics model handling which allows for simulations in the framework of
      the Standard Model, the MSSM, and some ADD model of extra dimensions,
      plus interfaces to some spectra generators (Hdecay and Isasusy) are
      implemented;
\item beam spectra handling to allow for treatment of Laser backscattering,
      Beamstrahlung, Weizsaecker-Williams-type processes, etc.;
\item a large set of PDF's that is easy to extend, but at the moment, the
      following sets are available: LHAPDF, MRST99 ({\tt C++}-version),
      CTEQ6 (Fortran version outside LHAPDF) as well as structure
      function for leptons;
\item a powerful matrix element generator ({\tt AMEGIC++}) as well as
      a - quite limited set - of simple $2\to 2$ processes in analytical
      form, both can be integrated with the full machinery of
      multi-channel integration;
\item their merging with the parton shower through the CKKW method,
      implemented for arbitrary processes;
\item a simple parton shower module ({\tt APACIC++});
\item an interface to the Lund-string hadronization of {\tt Pythia 6.163}
      and the subsequent hadron decays by the same program;
\item interfaces to HepEvt and HepMC as well as some simple
      analysis routines based on ROOT.
\end{itemize}

\subsection{Specialised Initial and Final State Radiation Programs}

\programAbstract{APACIC++}{F.~Krauss}
\programInfo{Tanju Gleisberg, Frank Krauss, Andreas Sch{\"a}licke,
Steffen Schumann, Jan Winter}%
{Ref.~\cite{Kuhn:2000dk} is the {\tt APACIC++} manual for version 1.0
(a manual for version 2.0 is forthcoming).}%
{ }
\programVersion{{\tt APACIC++} 2.0}

{\tt APACIC++} ({\tt A PA}rton {\tt C}ascade {\tt I}n {\tt C++})
is the parton shower module of the the new event generator {\tt SHERPA}
({\tt S}imulation for {\tt H}igh {\tt E}nergy {\tt R}eactions of
{\tt PA}rticles). In its original version (1.0) it carried much of the
functionality that has now migrated to the new framework. Specifically,
it was responsible not only for the multiple emission of partons
through the shower, but also for the merging with the matrix elements,
the interface to hadronization and hadron decays, and for the
overall event generation methods. Apart from the showering, all these
tasks have now been moved to {\tt SHERPA}, hence {\tt APACIC++}
cannot be used as a stand-alone program any longer. What remains in
version 2.0 of {\tt APACIC++} is only the parton shower in the initial
and final state.

This parton shower is done in a {\tt Pythia}-like fashion. In other words,
the ordering parameter of the radiation pattern is given by the virtual
mass of the partons; quantum coherence, i.e. angular ordering, is realized
only in approximate form by explicit vetoes on parton emissions with
rising opening angles. Differences to {\tt Pythia} are relatively minor, they
include:
\begin{itemize}
\item Full generic support for ME+PS merging, e.g. the possibility to apply vetoes
      in both the initial and final state parton shower on the emission
      of a parton that gives rise to an extra jet according to the $k_\perp$
      algorithm;
\item different treatment of heavy particles through modified splitting
      functions instead of cutting phase space;
\item different treatment of infrared cut-off of the parton shower;
\item abstract structure allowing for easy handling of splitting functions.
\end{itemize}

\programAbstract{Ariadne Colour Dipole Model}{L~L\"onnblad}
\programInfo{L.~L\"onnblad}%
{\cite{Lonnblad:1992tz}}%
{http://www.thep.lu.se/$\sim$leif/ariadne}
\programVersion{4.12}

The ARIADNE program\cite{Lonnblad:1992tz} implements the Colour
Dipole
Model\cite{Gustafson:1986db,Gustafson:1988rq,Andersson:1989gp,Andersson:1990ki}
for QCD cascades. It was initially developed to describe final-state
cascades in e$^+$e$^-$ annihilation, but has since been extended to
also describe collisions with incoming
hadrons\cite{Andersson:1989gp,Lonnblad:1995wk,Lonnblad:1996ex}. Here,
effects of initial-state radiation are described in terms of
final-state gluon radiation from colour dipoles produced in the hard
interaction, with special treatment (the so-called Soft Radiation
Model\cite{Andersson:1989gp}) of dipoles involving a hadron remnant.

The program has been very successful in describing data from LEP~1
and~2 and is also one of the few programs which are able to describe
the activity in the forward region in small-$x$ DIS events at HERA.
ARIADNE has not been compared extensively with data from
hadron--hadron colliders such as the Tevatron, although there is in
principle no problem to do so and, in particular in the forward
regions, the program should give different predictions as compared to
conventional parton shower based programs.  For e$^+$e$^-$
annihilation, ARIADNE includes a model for interfacing fixed-order
matrix elements with the dipole cascade\cite{Lonnblad:2001iq} which is
similar to the CKKW procedure\cite{Catani:2001cc}. Work is underway to
extend this model to also work for hadron collisions.

ARIADNE works as an add-on to PYTHIA, and a main program for PYTHIA
can be easily changed to use ARIADNE for the QCD shower, by simply
adding two function calls. The hard interactions, possible multiple
scatterings, hadronization and particle decays are then still handled
by PYTHIA. In principle this should work for any sub-process selected
in PYTHIA, although all of them have not been properly tested. In
particular, it has not been checked if the PYTHIA interface to the
Les Houches Accord Event record works together with ARIADNE.

ARIADNE is written in standard Fortran 77 and should be linked
together with a main program and PYTHIA.

\programAbstract{Photos}{Z.~Was}
\programInfo{E.Barberio, B. van Eijk, Z. Was}
{\cite{Barberio:1993qi,Barberio:1990ms}}%
{ http://wasm.home.cern.ch/wasm/goodies.html}
 
In the cascade decays of resonances, effects of QED bremsstrahlung
corrections need to be simulated.  Because of a multitude of decay
channels, the development of taylored solutions is not possible in every
case, and in fact is not necessary.  Photos can be used for
generation of bremsstrahlung corrections for the general case. The
precision of the generation may in some cases be limited, in general it
is not worse than the complete double bremsstrahlung in LL
approximation. The infrared limit of the distributions is also
correctly reproduced.  The action of the algorithm consists of
generating, with internally calculated probability, bremsstrahlung
photon(s), which are later added to the HEPEVT record. Kinematic
configurations are appropriately modified. Energy-momentum conservation
is assured. If difficulties arise relating to how the
event records are filled in by the host generator, the talk in
Ref.~\cite{kersevan:photos} may be useful.
Recently \cite{Golonka:2003xt}, technical documentation became available. 
Discussion of all recent improvements, in particular for case of $W$ 
decays is documented there.

\subsection{Programs for Diffractive Collisions}

\programAbstract{PHOJET}{R.~Engel}
\programInfo{Ralph Engel, Johannes Ranft, Stefan Roesler}
{\cite{Engel:1994vs,Engel:1995yd}}
{http://www-ik.fzk.de/$\sim$engel/phojet.html}
\programVersion{1.12}

The event generator PHOJET was developed for detailed modeling of
minimum bias events with a realistic superposition of various
types of diffractive and non-diffractive particle production processes.
The ideas and methods implemented in the program are based 
mainly on the Dual Parton Model (DPM) \cite{Capella:yb} and 
Quark-Gluon Strings Model \cite{Kaidalov86a}. 
The event generator is
formulated as a two-component model by distinguishing soft and hard 
components of multiparticle production, which are
combined in a sophisticated unitarization procedure
\cite{Engel:1994vs,Engel:1995yd,Aurenche:1991pk}.

PHOJET can be used to simulate
hadronic multiparticle production at high energies for hadron-hadron,
photon-hadron, and photon-photon interactions (hadron = proton,
antiproton, neutron, or pion). The generator includes the photon flux
simulation for photon-hadron and photon-photon processes in
lepton-lepton, lepton-hadron, and heavy ion-heavy ion collisions 
\cite{Engel:1996aa}. In
addition, various photon flux spectra of relevance to planned linear
colliders are implemented (bremsstrahlung, beamstrahlung,
laser-backscattering).

\programsubsection{Subprocesses}

All leading order matrix elements for scattering processes of quarks,
gluons and photons into light quarks and gluons are implemented.
By construction hard and semi-hard processes are not only simulated for
non-diffractive interactions but also for
single and double diffraction dissociation and central diffraction
(double pomeron exchange) \cite{Engel:1995sb,Bopp:1998rc}.

Up to now processes involving heavy quarks and $W$ and $Z$ vector 
bosons are not available in the code.

\programsubsection{Parton showers, hadronization and decay}

Initial state parton showers are simulated by a backward evolution
algorithm that uses parton density functions as external input
\cite{Sjostrand:1985xi} and includes some coherence effects by imposing angular
ordering. In the case of photons the anomalous term in the parton density
evolution equations is taken into account \cite{Drees:1993cm}.

Final state parton showers are generated by PYTHIA
\cite{Sjostrand:2000wi}, which is used to
handle string fragmentation, hadronization and resonance decays. A
number of spin/polarization-dependent decays are implemented separately
($\rho$, $\omega$ and $\phi$ production in photon diffraction
dissociation).

\programsubsection{Underlying event}

Soft and hard processes are treated in a unified way, applying a transverse
momentum cutoff to separate the two components of the model.
In general, PHOJET predicts multiple soft and hard interactions 
in one high-energy event.
Employing the optical theorem, Regge phenomenology is used to
parametrize various partial cross sections according to string and color
flow topologies.
The structure of the different event classes, including the soft
underlying particle production in events with hard interactions,
is thus predicted by the model
parameters that are found by fits to total, elastic, and diffractive
cross sections.

\programAbstract{POMWIG}{B.~Cox}
\programInfo{Brian Cox and Jeff Forshaw}%
{\cite{Cox:2000jt}}%
{http://www.pomwig.com}
 
POMWIG  \cite{Cox:2000jt} is a simple modification to the HERWIG Monte Carlo 
generator which allows the simulation of diffractive collisions. In proton - 
proton (or anti-proton) collisions,  both single and double diffractive 
collisions (sometimes known as 'double pomeron exchange') are implemented. In 
electron - proton collisions, the diffrative DIS process is implemented. In 
both cases, pomeron and reggeon exchange processes are generated seperately. 
By default, the pomeron and reggeon structure functions and flux factors are 
those measured by the H1 Collaboration \cite{Adloff:1997mi}, although POMWIG 
allows the user to implement new structure functions and flux factors in a 
simple way. 

POMWIG will run on any system that runs HERWIG (all versions from 5.9 onwards 
have been fully tested). Once the POMWIG routines have been added, HERWIG 
will function normally except for the generation of resolved  photoproduction 
events in electron - proton collisions (since it is these HERWIG routines 
which are modified in order to run POMWIG).

\subsection{Specialised Decay Programs}

\programAbstract{EVTGEN}{A.~Ryd}
\programInfo{David Lange and Anders Ryd}
{\cite{Lange:uf}}
{http://www.slac.stanford.edu/$\sim$lange/EvtGen}

The EvtGen package is a Monte Carlo program of resonance decays, focused on
the physics processes relavant to B meson decays.  The framework
includes tools needed to handle sequential decays and to correctly
simulate angular distributions, including their correlations.  Individual
physics processes are implemented in modules that allow users to
build complicated decay chains from simple pieces.  Each module
calculates decay amplitudes, used by the framework to generate the
correct kinematic distributions.  EvtGen provides
implementations of many detailed decays, including a variety of
semileptonic decay models and time dependent $CP$ asymmetries in
neutral B meson decays, as well as a
decay table for simulation of generic B decays.

EvtGen is written primarily in C++, with some legacy fortran
code.  EvtGen was mainly developed on the Linux platform,
but has been used on other platforms as well.
EvtGen interfaces to the PHOTOS package
for generation of final state radiation and to PYTHIA.
PYTHIA is used both for its hadronization capabilities as well as
to fill out the unknown component of the B meson decay table
via inclusive generic decays.  HEPEVT is used to interface
to both of these packages.

A hypernews forum is available at: \\
\href{http://www-babar2.slac.stanford.edu:5090/HyperNews/EvtGen/index/index.html}{
http://www-babar2.slac.stanford.edu:5090/HyperNews/EvtGen/index/index.html}.

\programAbstract{Tauola}{Z.~Was}
\programInfo{R. Decker, S. Jadach, M. Jezabek, J.H.Kuhn, Z. Was}
{\cite{Jadach:1993hs,Jezabek:1992qp,Jezabek:1991qp,Jadach:1990mz}}%
{http://wasm.home.cern.ch/wasm/goodies.html}

Tauloa is a Fortran 77 package used for generation of tau lepton decays including
spin polarization.  For each decay mode there is:
\begin{itemize}
 \item an individual phase space generator (with no approximation used);
 \item a part describing weak current: including first order QED
   corrections for leptonic decays and the possibility to admixture some
   non standard interactions, tau neutrino mass and free choice of
   vector and axialvector couplings of tau to virtual W state;
 \item a part describing hadronic current with several choices available,
   some of them are supported/distributed from Tauola web page, but
   nonetheless require individual referencing (available from the
   program printout);
 \item a part responsable for the choice of the decay mode
and overall administration, as well as for writing the generated decay
into HEPEVT record.

\end{itemize}

For more details look into the references, but also recent
transparencies of the MC4LHC workshop~\cite{Was:mc4lhc_tauola} or review
talk at the last tau conference, Ref.~\cite{Was:2002pu}.

A universal interface for Tauola to the HEPEVT event record is
provided in the Tauola Interface program, 
Ref.~\cite{Was:2002gv,Pierzchala:2001gc,Golonka:2003xt}.
Refer to \cite{Golonka:2003xt} for technical documentation.
Discussion of all recent improvements, in particular TAUOLA universal 
interface, is documented there.

A program called MC-Tester \cite{Golonka:2002rz} (Authors P. Golonka,
T. Pierzchala, Z. Was,\\
{\tt http://cern.ch/MC-TESTER })
is also available. This package, written in C++ and interfaced to
Fortran 77/90, is developed for tests of decay packages. The idea is
to have a quick way of comparing two packages for the decay of a
particle e.g. `X'. The algorithm searches over the input event records
(HEPEVT, PYJETS, LUJETS and some C++ records may be used as input) and
whenever a new decay of `X' is found, a list of the decay modes is
extended (classified on the basis of the decay products) and
histograms of all invariant masses are initialised on the first
occurence and later filled in. The data from two distinct runs of
MC-Tester with two decay packages can be later compared within a
MC-Tester analysis run, independently of the programming language or
event records used by the compared generators.
Since Les Houches workshop, MC-tester found a lot of applications 
outside decay libraries. In particular, it was found to uesful for 
comparisons of matrix element generators during  LC and LHC workshops 
and also in studies of software of ATLAS collaboration
(see \href{http://cern.ch/MC-TESTER/talks.html}).

\section[Resummation]{RESUMMATION~\protect\footnote{Contributed by:
E.~Laenen.}}\label{s_resummation}
\label{sec:resumm-what-that}

In this section we shall briefly discuss a different approach, with
respect to that of \SHG's, to the computations relevant to the 
phase-space regions where the cross sections typically peak.
We shall do so by answering the question: {\it Resummation, what is that?} 

Readers of this guidebook will readily answer that it refers to any effort at
summing some terms in a quantity's perturbative series to all orders.
This basic understanding, while correct, is not enough however to
really participate in, or perhaps fully appreciate, discussions
involving the merit of resummation in phenomenological issues.  To
this end it is profitable to know at least what the words ``some'',
``terms'', ``quantity'', and ``summing'' mean in the sentence above.
This brief section will, therefore, try to clarify these concepts
somewhat in the hope that such discussions might become more rewarding
for the reader. The text below leaves the quantum field theory
unspecified, but we have of course QCD in mind.

\textit{Quantity}: This is often an observable such as a
(differential) cross section, a decay rate, or a derived quantity like a
structure function. It might also be a more theoretical
quantity like a form factor, a parton distribution or fragmentation
function; in general it may be any quantity having a perturbative
expansion.

\textit{Terms}: Let $O$ be such a quantity with the (schematic)
perturbative expansion
\begin{equation}
  \label{eq:1}
  O_{PT} = f_{00} + \alpha (c_{12}L^2 + c_{11}L + f_{10}) + 
  \alpha^2 (c_{24}L^4 + c_{23}L^3 + c_{22}L^2 +  \ldots + f_{20}) + \ldots
\end{equation}
where $\alpha$ is the coupling of the theory, also serving as
expansion parameter, $L$ is some logarithm, and the $f_{i0}$ represent
all terms not containing a power of $L$. Our discussion here focuses
on the case with at most two extra powers of $L$ per order, associated
with an extra soft and/or collinear emission of a particle.  The
quantity $O$ determines what $L$ is the logarithm of: for a thrust
($T$) distribution $L = \ln(1-T)$, for $d\sigma(p\bar{p}\rightarrow
Z+X)/dp_T^Z$ $L = \ln(M_Z/p_T^Z)$. Note that $L$ does not
\textit{have} to be the logarithm of a measured variable but can also
be a function of unobservable partonic momenta to be integrated over, 
e.g. for inclusive heavy quark hadroproduction $L = \ln(1-4m^2/x_1 x_2
S)$ where $x_1,x_2$ are partonic momentum fractions.  When $L$ is
numerically large, so that even with small $\alpha$ the convergent
behaviour of the series is endangered, resumming the problematic terms
might remedy this and thereby extend the theory's predictive power to
the situation where $L$ is large.

\textit{Summing}: The (schematic) resummed form of $O$ may, in all
known cases, be written as
\begin{equation}
  \label{eq:2}
  O_{res} = \exp\left[L g_1(\alpha L) + g_2(\alpha L)+\ldots  \right]
 \left(f'_{00} + \alpha f'_{10} + \ldots\right)
\end{equation}
where $g_{1,2,\ldots}$ are known functions. Although even a sketch of
a derivation of such an expression is beyond the scope of this
section, we can discuss some of $O_{res}$'s features. First, the
residual series $\sum f'_{i0} \alpha^i$ is without logs and therefore
better-behaved.  The dependence on the logarithm has moved into the
exponent, which is now a series in $\alpha$, and under analytical
control. This is the main merit of resummation. Second, notice that the
resummed form contains an exponential, which reflects roughly the
Poisson statistics of independent emissions. Third, due to technical
reasons the $L$ in question in (\ref{eq:2}) is most often not the log
of the original variable (say, $p_T^Z$), but of a conjugate variable
(impact parameter $b$) resulting from a Fourier or other integral
transform.  An expression like (\ref{eq:2}) may be evaluated
numerically and used phenomenologically, involving of course the
appropriate inverse transform, but it should be mentioned that this is
not always an unambiguous procedure, in particular for QCD; the
all-order resummation can introduce infrared singular behaviour into
$O_{res}$ that is not present in finite order computations.
Therefore, a resummed result must, in such cases, be specified
together with a prescription how to handle this singular behaviour
numerically.

\textit{Some}: Specifying the theoretical accuracy for a perturbative
series such as eq.~(\ref{eq:1}) involves stating whether only the
leading order (LO) term has been kept, or also the next-to leading
${\cal O}(\alpha)$ (NLO) term, etc.  The analogue for the resummed
form (\ref{eq:2}) involves stating whether only $g_1$ is kept (leading
logarithmic (LL) accuracy) or also $g_2$ (next-to-leading
logarithmic (NLL)) is kept, etc.  Note that an increase in the
logarithmic accuracy must go along with including, without double
counting, more terms in the $\sum_i f'_{i0}\alpha^i$ series.  This is
called matching. Just as one may parametrically and systematically
increase the accuracy of the perturbative approximation (\ref{eq:1})
by including ever higher order terms, one may do so for the resummed
expression
by including ever more terms in the exponent, together with appropriate 
matching.\\

To summarise, a resummed quantity is, besides the fact that some of the 
terms in its perturbative expansions have been summed to all orders: 
characterised by the logarithm at hand, a statement of accuracy like
LL, NLL etc, and (possibly) a prescription to handle ambiguities.\\

In many cases, the authors of an observable's resummation 
write a computer code to study its effects numerically. Such 
codes are typically observable-specific and are often 
not written with a general user in mind (with some exceptions).
How, then, does resummation happen in event generators that are
purposely not observable-specific?

Recall that the simulation of an event by a particular generator
involves various stages: the hard subprocess, initial state showering, 
final state showering, and hadronization. Of these, the first three are
each described by perturbative physics involving interacting quarks,
gluons and other quanta. The hard subprocess is in all event
generators described by a leading order or next-to-leading order
matrix element. The shower algorithms, on the other hand, generate many
partons per event and include higher order contributions, because each 
parton generation requires at least one power of the coupling, 
in any Monte Carlo prediction of an observable. The algorithms are such
that in general the leading logarithms, whatever they may be, in the Monte 
Carlo prediction of the observable are correctly generated in this way, in 
addition to some, but not all, of the N$^k$LL, for any $k$. 
In essence, leading logarithms are process independent, whereas logarithms
beyond the leading ones usually are less so. But the reader should keep 
in mind that, while the logarithmic accuracy obtained
by a Monte Carlo resummation of an observable is almost always
less than in a dedicated study, the Monte Carlo can easily simulate 
acceptance cuts etc. Clearly, there is still lots of room for 
improvement in bringing these two descriptions, one analytical and
one by Monte Carlo, closer together. 

Finally, resummed calculations of observables are, in general, closely
linked with their power corrections. Some of the latter involve
hadronization effects. While there has been recent progress in 
understanding their connection more precisely for event shape observables
in $e^+e^-$ collisions and DIS, this is also an area where much remains to 
be understood.

\section[Combining Matrix Elements with Showering]
{COMBINING MATRIX ELEMENTS WITH SHOWERING~\protect\footnote{Contributed by: 
M.~Dobbs, S.~Frixione.}}\label{sect:MEwPS}

We have discussed at length the virtues of SHG's and fixed-order
(either tree-level multi-leg or NLO) predictions. The parton shower 
is most effective when the extra emissions are soft or collinear, 
which chiefly contribute to the peak of the cross section, whereas 
the matrix element prescription excels in the complementary region
(typically a high-$\pt$ tail). Ideally, one would like to model 
Nature with a program that knows both techniques and can use them 
simultaneously.

One common pitfall new users of SHG codes fall into is the combination
of a process like $q\bar{q} \rightarrow Z^0$, with its higher order
companion process $q\bar{q} \rightarrow Z^0g$. Events from these two
processes should never be blindly combined, since a fraction of the
latter events are already included in the former process via gluon
radiation in the parton shower. Combining the two processes without
special procedures amounts to {\em double counting} some portion of
phase space. However, using the first process alone is also 
unsatisfactory, because the parton shower does a poor job in modelling
the region of high transverse momentum of the $Z^0$.

This issue has been addressed in HERWIG and PYTHIA with {\em matrix element
corrections}. These can be implemented either as a strict partition of phase
space between the two processes, or as an event reweighting (re-evaluation of
the event probability using the matrix element) using the higher order tree
level matrix element for the related process. In either case the effect is the
same: the event shapes are dominated by the parton shower in the low-$\pt$
region, the shapes are NLO-like in the high-$\pt$ region, and the total cross
section remains leading order (i.e.\ for our example the total cross section
will be the same as that for $q\bar{q} \rightarrow Z^0$). The trouble with
matrix element corrections is that they can be applied only in a very
limited number of cases, which are relatively simple in terms of 
radiation patterns and colour connections. Furthermore, only one extra
emission can be treated with respect to the underlying hard subprocess.

New physics signals will likely be detected through multi-jet channels
(up to the order of ten jets), since heavy, fastly decaying particles
are expected to be formed. Standard SHG's such as HERWIG and PYTHIA,
regardless of the presence of matrix element corrections, perform
particularly badly for these observables. The obvious way out is that
of dressing the many hard partons available from a tree-level matrix
element generator (see sect.~\ref{sect:TLMEG}) with the extra emissions
provided by a shower mechanism. However, as pointed out in 
sect.~\ref{sect:TLMEG}, this procedure is not completely safe and
a dependence can arise of physical observables upon {\em unphysical} 
parton cuts (which we symbolically denote as $y_{cut}$). Typically, this 
dependence is of leading log nature (i.e., $\as^k \log^{2k}y_{cut}$).

A solution to this problem has been proposed in ref.~\cite{Catani:2001cc}
(referred to as CKKW, after the names of the authors). The phase space
of $n$ partons is partitioned, using the parameter $y_{cut}$, into two
regions, which can be called parton shower dominated and matrix
element dominated. In the former region, the hard kinematics is that 
relevant to $n-1$ partons; these kinematics act as an initial condition for 
a {\em vetoed shower}, where the veto basically prevents the shower from 
populating the latter region. In the matrix element dominated region the
hard kinematics is that of $n$ partons. In both regions, the matrix elements
are reweighted with a suitable combination of the Sudakov form factors
entering the shower algorithm. It is clear that, in order to be 
internally consistent, matrix elements must be available for any value
of $n$. In practice, $n\le 5$ is a good approximation of $n<\infty$.
Using the CKKW prescription, the dependence of the physical observables
upon $y_{cut}$ is reduced from leading to next-to-next-to-leading log (i.e., 
$\as^k \log^{2k-2}y_{cut}$), plus terms suppressed by powers of $y_{cut}$.
Although the original CKKW proposal concerned $e^+e^-$ collisions, an 
extension to hadronic collisions has been presented~\cite{Krauss:2002up}, 
and practical implementations have been achieved in HERWIG, PYTHIA, and 
SHERPA (see sect.~\ref{s_shg_programs}). There is considerable freedom
in the implementation of the CKKW prescription in the case of hadronic
collisions.  This freedom is used to tune (some of) the SHG's parameters in
order to reduce as much as possible the $y_{cut}$ dependence, which
typically manifests itself in the form of discontinuities in the
derivative of the physical spectra. We note that the complete 
independence of $y_{cut}$ cannot be achieved; this would be possible
only by including all diagrams (i.e., also the virtual ones) contributing
to a given order in $\as$. 

In the last couple of years, the problem of including in SHG's the 
complete higher-order corrections to matrix elements has received
considerable attention. Given the situation of the fixed-order computations
described in sect.~\ref{sect:NLOcomp}, the only case which could be
studied in practice is that of the NLO matrix elements. Remarkably, 
a few proposals have passed the stage of theoretical exercises and
made it to the implementation step. The corresponding codes are 
presented below and readers interested in the technicalities of
the formalisms are urged to check the original papers.

\subsection{Programs using NLO Matrix Elements with Showering} 

\programAbstract{grcNLO (GRACE NLO with Parton Shower)}{Y.~Kurihara}
\programInfo{Y. Kurihara, J. Fujimoto, T. Ishikawa, K. Kato,
S. Kawabata, T. Munehisa, H. Tanaka.}{\cite{Kurihara:2002ne}}%
{http://research.kek.jp/people/kurihara/}
\programVersion{Program is not yet available.}

A new method to construct event-generators based on next-to-leading 
order QCD matrix-elements and leading-logarithmic parton showers is 
proposed.  Matrix elements of loop diagrams as well as tree 
level can be generated using an automatic system. A soft/collinear
singularity is treated using a leading-log subtraction method.
Higher order re-summation of the soft/collinear correction by the parton 
shower method is combined with the NLO matrix-element without 
any double-counting in this method.

\programAbstract{MC@NLO}{S.~Frixione}
\programInfo{S.~Frixione, B.R.~Webber}{
\cite{Frixione:2002ik,Frixione:2003ei}}
{http://www.hep.phy.cam.ac.uk/theory/webber/MCatNLO/}
\programVersion{2.3}

The MC@NLO event generator includes the full next-to-leading order 
QCD corrections in the computation of hard subprocesses. It is based 
on the formalism presented in refs.~\cite{Frixione:2002ik,Frixione:2003ei}.
In the current version, the package includes hadronic collisions, 
with the production of the following final
states: $W^+W^-$, $W^\pm Z$, $ZZ$, $b\bar{b}$, $t\bar{t}$,
$H^0$, $W^\pm$, $Z$, $\gamma^*$, $l_i\bar{l}_j$, with the latter lepton
pair originating from an off-shell $W^\pm$, $Z$ or $\gamma$ (the $Z/\gamma$
interference is included). Mass effects are always included; spin 
correlations for the decay products are included except in the cases of
vector boson pair and $t\bar{t}$ production.

Incorporating the NLO matrix elements provides a better prediction of
the rates while improving the description of the first hard parton
emission. As with any other parton shower based Monte Carlo, MC@NLO is
capable of giving a sensible description of multiple soft/collinear
emissions. For the same reason, and at variance with usual NLO
programs, propagation through the shower and subsequent hadronization
gives a final state description at the hadron level. One feature of
MC@NLO as opposed to standard MC's is the presence of negative
weights. Therefore in unweighted event generation MC@NLO produces unit
weight events with a fraction (typically 15\%) having weight -1.
The unweighting efficiency is of the order of 40\% or higher for all
the processes considered. Weighted event generation has not been 
included, but there is no principle reason which prevents it.

The program is provided as a standalone package written in Fortran77, 
downloadable from the web site given above. It writes an event file 
which is read by a general purpose showering and hadronization code 
using the Les Houches Interface Standard~\cite{Boos:2001cv}.
Although the MC@NLO formalism is general, in the current version such
showering and hadronization code {\em must} be HERWIG (version 6.5 
or newer). Bash scripts and a Makefile are provided to run the 
code in a way similar to standard HERWIG (in fact, the same analysis
routines can be used). The code has been tested on various operating systems:
Linux, Sun Unix, Digital Unix running on Alpha's, Mac OSX.
The package includes a self-contained library of parton densities
(updated to including the CTEQ6 and MRST2002 families), and an interface
to PDFLIB.

\programAbstract{Phase Space Veto}{M.~Dobbs}
\programInfo{M.~Dobbs}{\cite{Dobbs:2001dq}}{none}
\programVersion{Proof-of-concept only. 
Not currently supported. Contact author.}

The phase space veto is a method for organising next-to-leading order
QCD calculations using a veto which enforces the cancellations between
virtual and real emission diagrams, leaving a region of phase space
where the Parton Shower method can be employed. 
Essentially the method partitions phase space and uses {\it either}
the NLO matrix element {\it or} the
parton shower method in each region. In this manner no region is
counted twice, but in the (soft and collinear) domain of the parton
shower, the event shapes are not precisely accurate to NLO. The
advantage of this technique over other methods is that samples of
truely unweighted events can be produced (there are no negative
weights and no events which must be used to cancel other events). The
total cross section from this method is precisely NLO.

The method employs phase space slicing with the slicing parameter
determined dynamically event-by-event. The output can be interfaced to
general purpose showering and hadronization programs to obtain
complete event descriptions. 
Only one proof-of-concept process, $\ppORppbar\rightarrow Z + X$, is
implemented. This process is interfaced to the \PYTHIA\
shower and hadronization package. The program is written in C++ using
modern object-oriented techniques.

\section[Conclusions]{CONCLUSIONS~\protect\footnote{Contributed by: 
the editors.}}\label{s_conclusions}
A survey of Monte Carlo programs for the simulation of hadron collider
events has been presented with the aim of making the programs more
accessible to a new user.

The reader familiar with Monte Carlo codes employed during the last
decade will notice a trend in modern simulation programs. They are
becoming more modularised, with authors specialising their codes to a
focused aspect of the event simulation. The user has the luxury of
choosing different tools for different aspects of the event---and the
responsibility to understand the limitations and caveats of each
tool's use.
The community is moving towards a time when each aspect of
the event simulation (hard subprocess, parton shower, etc.) can be
interchangeably simulated with different programs, allowing for the
cross checking of results and an estimate of the systematic errors
associated with each aspect. Breakthroughs in the merging of seemingly
distinct techniques (the Parton Shower with NLO matrix elements) have
been achieved, and will be of ever greater importance as colliders
move towards higher energy.

A few brave program authors are embracing modern software programming
languages, either by rewriting existing codes or beginning new
projects using object-oriented languages such as C++. This fits very
well with the current trends for detector simulation (such as the
complete rewriting of GEANT in C++~\cite{Agostinelli:2002hh}), and the
trend for experimental collaborations, who are overwhelmingly choosing
C++ for their experiment software.

With modern modular Monte Carlo simulation tools, the complexity of
the event generation chain is approaching that of a complicated
detector subsystem. Given that the development of these tools is
struggling along with a fraction of the resources and funding
typically allocated to an experiment's software, it's amazing what has
been achieved in this field.

\section[Acknowledgments]{ACKNOWLEDGMENTS}\label{s_acknowledgments}

The Les Houches ``Physics at TeV Colliders'' workshop is a stimulating
environment. We thank the organisers of the workshop for putting
together a first class event. We thank the participants for providing
a high level atmosphere and many stimulating discussions.

Many Monte Carlo authors contributed to this document, including
several who did not attend the workshop. The editors thank them all
for taking time to summarise their programs in this guidebook.

\small{The work of M.~Dobbs was supported in part by the Director, Office 
  of Science, Office of Basic Energy Sciences, of the U.S. Department of Energy
  under Contract No. DE-AC03-76SF00098.}


\end{document}